\documentclass[journal, 10pt, twocolumn]{IEEEtran}
\usepackage{graphicx}
\usepackage{hyperref}
\usepackage{amssymb}
\usepackage{amsmath}
\usepackage{bm}
\usepackage{color}
\usepackage{enumerate}
\usepackage{algorithm}
\usepackage{algorithmic}
\usepackage{color}

\newtheorem{thm}{\textbf{Theorem}}[section]
\newtheorem{prop}{\textbf{Proposition}}[section]
\newtheorem{lem}{\textbf{Lemma}}[section]

\usepackage[compress,sort,comma,numbers]{natbib}

\newcounter{temp}

\begin{document}

\title{Incentivized Campaigning in Social Networks }

\author{\IEEEauthorblockN{Bhushan Kotnis, Albert Sunny, Joy Kuri}
\thanks{Department of Electronic Systems Engineering, 
Indian Institute of Science, Bangalore, India.  
Email: \{bkotnis,salbert,kuri\}@dese.iisc.ernet.in}
}

\maketitle

\begin{abstract}
Campaigners, advertisers and activists are increasingly turning to social recommendation mechanisms, provided by social media, for promoting their products, services, brands and even ideas. However, many times, such social network based campaigns perform poorly in practice because the intensity of the recommendations drastically reduces beyond a few hops from the source. A natural strategy for maintaining the intensity is to provide incentives. In this paper, we address the problem of minimizing the cost  incurred by the campaigner for incentivizing a fraction of individuals in the social network, while ensuring that the campaign message reaches a given expected fraction of individuals. We also address the dual problem  of maximizing the campaign penetration for a resource constrained campaigner. To help us understand and solve the above mentioned problems, we use percolation theory to formally state them as optimization problems. These problems are not amenable to traditional approaches because of a fixed point equation that needs to be solved numerically. However, we use results from reliability theory to establish some key properties of the fixed point, which in turn enables us to solve these problems using algorithms that are linearithmic in maximum node degree. Furthermore, we evaluate the efficacy of the analytical solution by performing simulations on real world networks.  
\end{abstract}


\section{Introduction}
On-line social networking services have enabled advertisers, campaigners and activists to reach millions of individuals. In particular, the ability to recommend or share web articles  \cite{Leskovec2009}, videos \cite{Leskovec2006}, and other media can be harnessed by campaigners for disseminating information to a worldwide audience.  
While such social network based campaigns sound promising, due to the fact that ideas propagate only a few hops from their origins \cite{Dinh2012}, often, they are found to perform poorly in practice. 
Providing incentives for recommendations is a natural solution for increasing the hop count. 
For example, Dropbox, which offers cloud storage services, increased its customer base rapidly by offering incentives for social recommendations \cite{Houston2010}. 
 Although such referrals may increase the short term costs to the company by  generating a large number of registrations for the free service, in the long term, the free registrations pay off, since a significant portion of free users eventually migrate to the premium service. Also, to keep the cost down, they capped the referral payouts --- additional 500 MB  only for the first 28 referrals, i.e. the referral payout is capped at 14 GB. 

 An alternative mechanism to limiting the cost is to cap the number of incentivized individuals, i.e., instead of capping incentive payments, offer large referral rewards\footnote{For example the living social website gives $100\%$ cash-back on a purchased product if the customer persuades three others to buy the (same) product. Another famous example of uncapped referrals is Uber, and also the taxi service Lyft.} to a preselected set of individuals\footnote{{Instead of advertising the referral scheme, a randomized scheme can be advertised. For example, the advertisement could be ``those who register for the service will get a chance to win discounts for referring their friends."}}. This approach allows the campaigner to target individuals who are best suited to use and recommend the service to their co-workers or friends. 
 
 
\subsection{Related Work}
{In \cite{Kempe2003,Chen2009,Chen2010}, the authors consider nodes to be either active or inactive. Making an inactive node active --- not through network effects, but through direct intervention for kick-starting the campaign --- is termed as ``seeding.'' Assuming full knowledge of the network structure, i.e., the adjacency matrix, the question asked in \cite{Kempe2003,Chen2009,Chen2010} is; Given a constraint on the number of seeds, what is the optimal set of seed nodes that would maximize the reach of the campaign? However, in several real-world settings, we may just have access to the statistical properties of the network. Also, the authors in \cite{Kempe2003,Chen2009,Chen2010} do not consider the cost incurred due to incentivization. Incentivization is the process of providing incentives to nodes who are already active, to persuade their friends to sign up.  Incentivization happens throughout the campaign, whereas seeding, as discussed in \cite{Kempe2003}, happens only at the beginning. }

 A related problem involving the calculation of an optimal pricing strategy for products sold  to individuals in a social network was explored in \cite{Hartline2008,Arthur2009}. The authors in \cite{Hartline2008,Arthur2009} consider the problem of pricing a product and calculating the amount of cash-back (discount) that is provided to individuals as an incentive to evangelize the product. In this paper, we do not focus on optimal pricing, but rather focus on the size and cost of the campaign where individuals \emph{do not} incur a cost to register for the service.  The problems studied in this paper are more appropriate in settings where the service is free, 
 or is based on a freemium model. Furthermore, we only assume partial knowledge of the network and also incorporate constraints that ensure a given campaign penetration, whereas the algorithms proposed in \cite{Hartline2008,Arthur2009} assume full knowledge of the network structure and do not consider any constraint on the campaign size. 

The problem of computing the optimal referral payment mechanisms that maximize profit was studied in \cite{Lobel2014}, by modelling the referral process as a network game. The authors in \cite{Lobel2014} conclude that a combination of linear payment mechanism (linear in the number of referrals) and threshold payment mechanism (payment only when number of referral exceeds a threshold) approximates the optimal pricing scheme. In this paper, we focus on the set of nodes to be incentivized while assuming that a pricing scheme, which can be computed based on the results in \cite{Lobel2014}, is provided by the campaigner. Similar problems involving the computation of referral rewards in real time, for maximizing the campaign spread, were studied using the theory of optimal control in \cite{Karnik2012,Dayama2012,Kandhway2014,Kandhway2014a,Kandhway2014b}. 


\subsection{Our Contributions}
We consider a scheme where preselected incentivized individuals are presented with a reward when they register for the service, encouraging them to spread the news about the service to their friends. 
 The decision of whether to offer an incentive to an individual is precomputed based on the solution of an optimization problem. We use a variant of the linear threshold model \cite{Kempe2003} for modelling the campaign spreading process. For a given fraction of such incentivized individuals, we first compute the campaign size (expected fraction of registered individuals) using bootstrap percolation. We then use this quantity to formulate the following optimization problems: 1). minimize the cost for achieving a given expected fraction of registered individuals, and 2). maximize the expected fraction of registered individuals for a given cost budget. These optimization problems are not amenable to traditional approaches because of a fixed point equation that needs to be solved numerically. However, we use results from reliability theory to establish some key properties of the fixed point, which in turn enables us to solve these problems using algorithms that are linearithmic in the maximum node degree. Through extensive simulations, we also study the efficacy our incentivization scheme in real world networks.  

\begin{table}[h]
	\centering
	 {
	\caption{A summary of key notation \label{table:notation}}
	\begin{tabular}{|r|l|}
		\hline \textbf{Notation} &  \textbf{Description} \\ 
		\hline $p(k)$ & fraction of degree $k$ nodes \\ 
		\hline $k_{max}$ & maximum node degree of graph $\mathcal{G}$ \\
		\hline $p_{th}(m|k)$ & probability of a degree $k$ node having threshold $m$ \\
		\hline $\alpha_1 $ & activation probability of a type $1$ node\\
		\hline $\alpha_2 $ & activation probability of a type $2$ node\\
		\hline $\overline{d}$ &  mean node degree of graph $\mathcal{G}$  \\
		\hline $q$ & probability of encountering a type~$2$ node \\
		&  by traversing a randomly chosen link\\
		\hline & random variables that denote the number of type~2 \\
		$X_{k_2}$  & \emph{active neighbours} of a degree $k$ node, given that there \\
		& are $k_2$ type~2 neighbours\\
		\hline & random variables that denote the number of type~1 \\
		$Y_{k-k_2}$  & \emph{active neighbours} of a degree $k$ node, given that there \\
		& are $k-k_2$ type~1 neighbours\\
		\hline $u$ & probability of reaching a registered node by following \\
		& an arbitrary edge of the graph\\
		\hline $u_q$ & fixed point of Equation~\eqref{eq:u1_fixed_point} \\
		\hline $s(q)$ & expected fraction of registered nodes \\
		& at the end of the campaign\\
		\hline $s_k(q)$ & expected fraction of degree $k$ registered \\
		& nodes at the end of the campaign\\
		\hline $c_k$ & cost of incentivizing a degree $k$ node \\
		\hline $\phi(k)$ & probability of incentivizing a degree $k$ node \\
		\hline $\gamma$ & minimum expected fraction of registered nodes\\
		\hline $\overline{c}$ & maximum expected incentivization cost \\
		\hline
	\end{tabular} 	}
\end{table}

{The key notation used in this paper is summarized in Table~\ref{table:notation}}. The remainder of this paper is organized as follows. In Section~\ref{sec:model}, we present the campaign model. In Section~\ref{sec:compute_cascade_size}, using results from percolation theory, we compute the campaign size (expected proportion of registered individuals). Using this quantity,  we formulate and solve two relevant optimization problems in Sections~\ref{sec:cm} and \ref{sec:csm}. In Section~\ref{sec:num_eval}, we compare the analytical results with simulation performed on real-world networks. Finally, in Section~\ref{sec:conclusion}, we conclude the paper. 

\begin{figure}[t]
	\centering
	\includegraphics[scale=0.6]{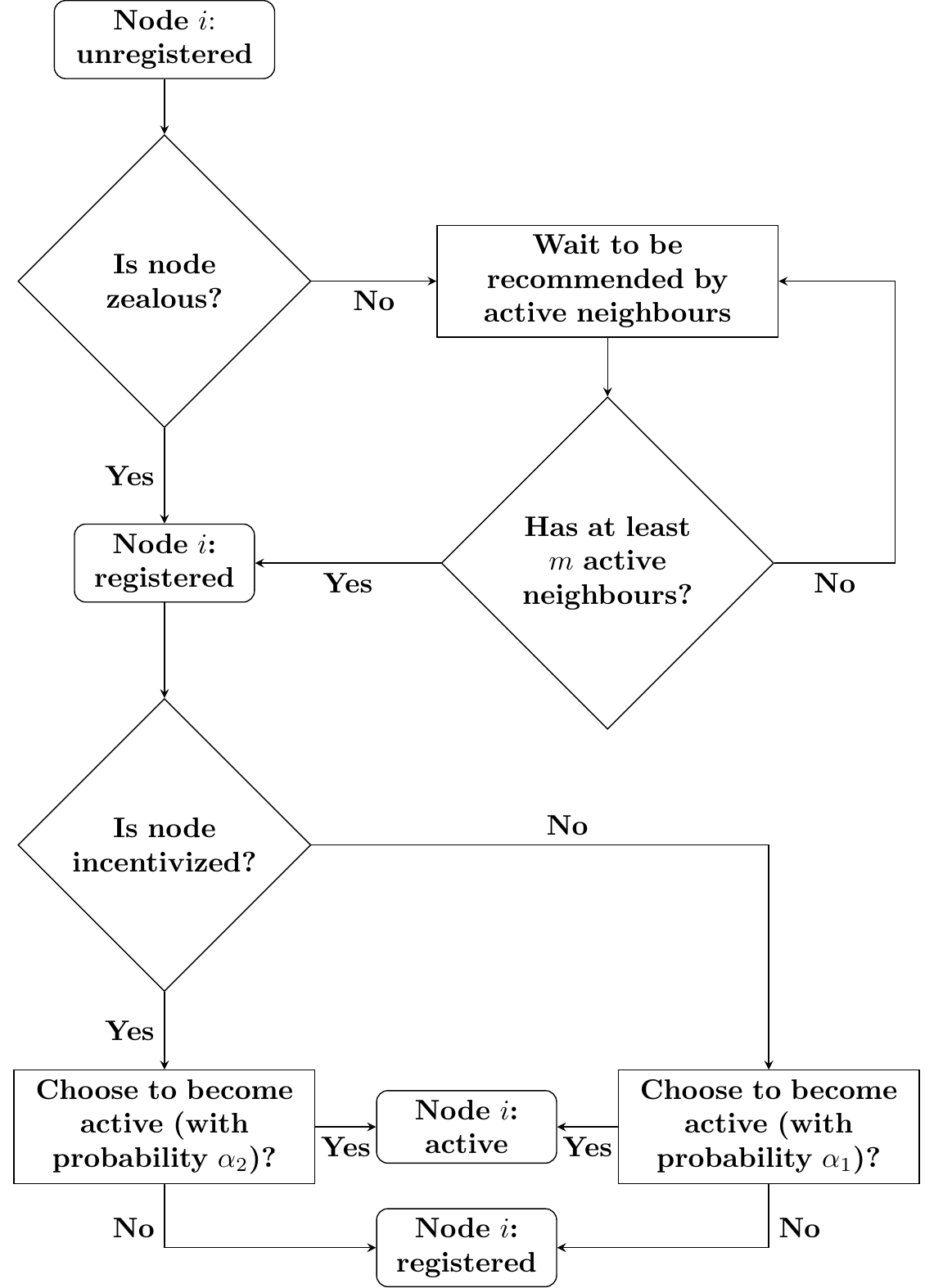}
	\caption{Flow chart denoting the various stages of a node; $m$ denotes the \emph{threshold} of node $i$, $\alpha_1$ and $\alpha_2$ are the activation probabilities of  \emph{incentivized} and \emph{non-incentivized} nodes, respectively. \label{fig:flow_chart}}
\end{figure}

\section{Model} \label{sec:model}
Consider a set of $\mathcal{N} = \{1, 2, \cdots, n\}$ individuals who are connected to one another through a social network. {For analytical tractability, as in \cite{Barrat2008,Baxter2010}, we consider this network to be arbitrary, connected\footnote{{The configuration model random graph is connected with high probability if and only if the minimum degree is greater than or equal to $3$.}}, locally tree-like and uncorrelated\footnote{{Uncorrelated networks are networks where the degree of a node is statistically independent of the degree of any other node in the network \cite{Dorogovtsev2010}.}}}. We represent this network as an undirected graph $\mathcal{G}(\mathcal{N}, \mathcal{L})$, where $\mathcal{N}$ and $\mathcal{L}$ represent the set of nodes and links of graph $\mathcal{G}$, respectively. An undirected edge $(a,b) \in \mathcal{L}$ if individuals $a$ and $b$ are neighbours in the underlying social network. {In most scenarios, full knowledge of the network structure may not be available to the campaigner. In such cases, the campaigner can obtain statistical properties of the network through data mining. One such property is the degree distribution.} 
Let $\{p(k), k \geq 1 \}$ be the degree distribution of graph $\mathcal{G}$.

 We consider a campaign on the network represented by graph $\mathcal{G}$ (see Fig.~\ref{fig:flow_chart} for the flow chart).
The nodes in graph $\mathcal{G}$ can be either in \emph{``active''}, \emph{``registered''} or \emph{``unregistered''} state. Once active, a node remains active. Active nodes are very spirited and express their strong support for the campaign by encouraging their neighbours to register, whereas nodes in the registered and unregistered states refrain from recommendations. For a node to become active, it must first show support for the campaign by registering itself. However, all registered nodes need not be active. 

The campaign starts with a set of \emph{zealous} individuals --- those who intrinsically desire the service. These nodes do not need any recommendation to register, and are registered for the campaign from the very start. If a zealous node becomes active, it will recommend the service to its neighbours. On the other hand, nodes that are not zealous, register only if the number of active neighbours exceeds a predefined threshold value (see Fig.~\ref{fig:flow_chart}). Let $p_{th}(m|k)$ denote the probability of a degree $k$ node having threshold $m$. {This is a generalization of the celebrated linear threshold model \cite{Kempe2003}} --- a model of choice for studying the dissemination and adoption of new products, technologies and ideas \cite{Granovetter1978,Macy1991,Mahajan1991,Berger2001}. We note that, by definition, $\sum_{k \geq 1} p(k) \cdot p_{th}(0|k)$ denotes the fraction of zealous nodes in the network\footnote{Campaigner can estimate the fraction of zealous nodes by conducting surveys on a reasonable sized demographic.} --- fraction of nodes that register without any recommendations. 

 
 A node that receives a reward when it registers for the service is more likely to tell its neighbours about the reward and encourage them to register. Therefore, we divide the nodes into two categories: the  `\emph{non-incentivized}' (type $1$)  and the `\emph{incentivized}' (type $2$).  The campaigner decides the fraction of incentivized nodes before the start of the campaign\footnote{Alternatively, whenever an individual registers, the campaigner can dynamically decide whether to incentivize this individual. This problem is similar to stochastic control problems, and is beyond the scope of this paper.}. However, these nodes become aware of the fact that they are incentivized only if they register for the service\footnote{{Incentives are assigned assuming knowledge of the exact degree of the registered node, because the exact node degree can be obtained after registering. For example, one can ask individuals to register through an on-line social network, and compute the exact degree from their contact list.}}. 
Since incentivized nodes are provided with incentives upon registration, they are more likely to be enthused, and will play an active part in the campaign than non-incentivized nodes. Let $\alpha_1 \in (0,1]$ and $\alpha_2 \in (0,1]$ be the probabilities of a non-incentivized (type $1$) and incentivized (type $2$) node becoming active, respectively.  Due to the presence of incentives, we assume that $\alpha_2>\alpha_1$. 


\begin{figure*}[!t]
\setcounter{temp}{\value{equation}}
\setcounter{equation}{5}
\begin{align} 
f(q, u) &=  \sum \nolimits_{k \geq 0} p_{ex}(k) \cdot  p_{th}(0|k+1)  + \sum \nolimits_{k \geq 1}  p_{ex}(k) \sum \nolimits_{m \geq 1} p_{th}(m|k+1) \sum \nolimits^{k}_{k_2=0} \hat{p}(k_2 | k) \cdot  P[X_{k_2} + Y_{k-k_2} \geq m ] \nonumber \\
&= \sum \nolimits_{k \geq 0} p_{ex}(k) \cdot  p_{th}(0|k+1)  + \sum \nolimits_{k \geq 1} p_{ex}(k) \sum \nolimits_{m \geq 1} p_{th}(m|k+1) \sum \nolimits^{k}_{k_2=0} {k \choose k_2} \cdot q^{k_2} \cdot (1-q)^{k - k_2} \cdot  \nonumber \\
& \hspace{40mm} \left(\sum \nolimits_{l + l^{'} \geq m} {k_2 \choose l} (\alpha_2u)^l (1-\alpha_2u)^{k_2-l} {k-k_2 \choose l^{'}} (\alpha_1u)^{l^{'}} (1-\alpha_1u)^{k-k_2-l^{'}} \right) \nonumber \\
g_k(q, u) &=   p_{th}(0|k)  + \sum \nolimits_{m \geq 1} p_{th}(m|k) \sum \nolimits^{k}_{k_2=0} \hat{p}(k_2 | k) \cdot  P[X_{k_2} + Y_{k-k_2} \geq m ] \nonumber \\  
& \hspace{-12mm} \textrm{where } X_{k_2} \sim Bin(k_2,\alpha_2u) \textrm{ and } Y_{k-k_2} \sim Bin(k-k_2,\alpha_1u) \nonumber
\end{align}
\setcounter{equation}{\value{temp}}
\hrulefill
\vspace*{4pt}
\end{figure*}

 \section{Computing the Cascade Size} \label{sec:compute_cascade_size}

Before presenting the problems, we need to first compute cascade size, i.e., the expected fraction of nodes that have registered at the end of the campaign. Let $p(k' | k)$ be the probability of encountering a degree $k'$ node by traversing a link from a degree $k$ node. 
 From \cite{Barrat2008}, we know that this conditional distribution, {for an uncorrelated network}, has the following form $p(k' | k) = \frac{k' \cdot p(k')}{ \overline{d}}$, where $\overline{d}$ denotes the mean degree of the graph $\mathcal{G}$.  Let $q(k)$ be the probability of encountering a type~$2$ node by traversing a randomly chosen link from a degree $k$ node. Then, we have
 \begin{align*}
 q(k) &= \sum \nolimits_{k^{'} \geq 1} \textrm{Pr}[\textrm{type~2 neigh.} | k^{'}] \cdot p(k^{'}| k) \nonumber \\
 &= \sum \nolimits_{k^{'} \geq 1} \phi(k^{'}) \cdot p(k^{'}| k)  = \frac{1}{\overline{d}}\sum \nolimits_{k^{'} \geq 1} k^{'}  \phi(k^{'})  p(k^{'}) 
 \end{align*}
 
 Since $q(k)$ is independent of $k$, let $q(k) = q \ \forall \ k \in \{1,2, \cdots ,k_{max}\}$. Here, $k_{max}$ is the maximum node degree. 
    
Since the network is locally tree-like, with some effort, it can be shown that the number of type~$2$ neighbours of a degree $k$ node is a \emph{binomial random variable} with the probability distribution function  $\hat{p}(k_2 | k) = {k \choose k_2} \cdot q^{k_2} \cdot (1-q)^{k - k_2}$.  Let $p_{ex}(k^{'})$ be the excess degree distribution, i.e., the degree distribution of a node encountered by following a randomly chosen link without counting that link. For an uncorrelated network, from \cite{Baxter2010}, we know that $p_{ex}(k^{'}) = \frac{(k^{'}+1) \cdot p(k^{'}+1)}{ \overline{d}}$.

Now, let $u$ denote the probability of reaching a registered node by following an arbitrary edge of the graph. First, let us consider an arbitrary node $j$ of degree $k$ and threshold $m$. Next, we compute the probability that node $j$ registers due to recommendations from its active neighbours, given that we have arrived at node $j$ by following an arbitrary link in the graph. For ease of presentation, we do not explicitly write the condition  ``following an arbitrary link.''
\begin{align}
&\hspace{0mm} P[j \textrm{ registers } | j \textrm{ is of degree } k, j \textrm{ has threshold } m] \nonumber \\
&\overset{(a)}{=}  {\textstyle \sum^k_{l = m}} P[j \textrm{ has } l \textrm{ active neighbours} | j \textrm{ is of degree } k] \nonumber \\
&= {\textstyle \sum^k_{l = m}} {\textstyle \sum \nolimits^k_{k_2=0}} P[k_2 \textrm{ type~2 neighbours} | j \textrm{ is of degree } k] \nonumber \\
&   \hspace{2mm} \cdot P[l \textrm{ active neigh.}| j \textrm{ is of degree } k \textrm{ and } k_2 \textrm{ type~2 neigh.} ] \Big) \nonumber \\
&= {\textstyle \sum^k_{l = m}} \left( {\textstyle \sum^{k}_{k_2=0}} \hat{p}(k_2 | k) \cdot  P[X_{k_2} + Y_{k-k_2} = l ]\right) \nonumber \\
&= {\textstyle \sum^{k}_{k_2=0}} \hat{p}(k_2 | k) \cdot  P[X_{k_2} + Y_{k-k_2} \geq m ] \label{eq:p_j_reg}
 \end{align}
 \normalsize
{where Equality~(a) follows because the events $\{j$ registers$\}$ and $\{j$ has $m$ or more active neighbours given that  $j$ has threshold $m\}$ are equivalent, and the events $\{j $ has $l$ active neighbours$\}$ and $\{j$ has threshold $m\}$ are independent of each other.}

In Equality~\eqref{eq:p_j_reg}, $X_{k_2}$ and $Y_{k-k_2}$ are random variables that denote the number of type~2 and type~1 \emph{active neighbours} of a degree $k$ node, given that there are $k_2$ type~2 and $k-k_2$ type~1 neighbours. With some effort, it can be shown that random variables $X_{k_2}$ and $Y_{k-k_2}$ are independent and have a binomial distribution with parameters $(k_2,\alpha_2 u)$ and $(k-k_2,\alpha_1 u)$, respectively. Here, $\alpha_1$ and $\alpha_2$ are the probability of a type~1 and type~2 node becoming active, respectively. {The independence occurs because the probability of incentivizing a node is independent of the degree of its neighbouring nodes. The binomial distribution arises here because the network is locally tree-like. }


From Equation~\eqref{eq:p_j_reg}, it is evident that the probability of a node registering is \emph{independent of its type}\footnote{A node registers either if it is zealous (this event is independent of its type), or through recommendations which depend on its neighbours and type of neighbours but not on its type.}. Therefore, the probability that by following an arbitrary link we can reach a  node that registers due to recommendations is given by
\begin{align}
&P[\textrm{node registers due to recommendation}] = \sum \nolimits_{k \geq 1} p_{ex}(k) \cdot \nonumber 	\\
&\hspace{5mm}  P[\textrm{node reg. due to recommendations} | \textrm{excess degree } k]  \nonumber 
\end{align}
 \begin{align}
&= \sum_{k \geq 1} p_{ex}(k) \hspace{-1.5mm} \sum_{m \geq 1}  \hspace{-1mm} P[\textrm{node registers, node has threshold } m | \nonumber \\
& \hspace{45mm} \textrm{node has excess degree } k] \nonumber \\
&\overset{(a)}{=} \sum_{k \geq 1} p_{ex}(k) \hspace{-1.5mm} \sum_{m \geq 1}  \hspace{-1mm} p_{th}(m|k+1) \cdot P[\textrm{node registers} | \nonumber \\
& \hspace{12mm} \textrm{node has excess degree } k,\textrm{ node has threshold } m] \nonumber \\
& =\sum_{k \geq 1} p_{ex}(k) \sum_{m \geq 1} p_{th}(m|k+1) \cdot \nonumber \\
& \hspace{3mm} \sum^{k}_{k_2=0} {k \choose k_2} \cdot q^{k_2} \cdot (1-q)^{k - k_2} \cdot  P[X_{k_2} + Y_{k-k_2} \geq m ] \label{eq:node_reg}
\end{align}

{On the RHS of Equation~\eqref{eq:node_reg}, we use the excess degree distribution because we discount the link that we followed to arrive at the node. In Equality~(a), we use $p_{th}(m|k+1)$ because if we include the link on which we arrived, a node of excess degree $k$ will have $k+1$ links}. Arguing along the lines of \cite{Baxter2010}, we can conclude that $u$ has to satisfy the following self consistency equation
\begin{align}
& u  = P[\textrm{node is zealous}] \nonumber \\
 & \hspace{20mm}  + P[\textrm{node reg. due to recommendations}] \nonumber \\
&= \sum \nolimits_{k \geq 0} p_{ex}(k)  p_{th}(0|k+1) \nonumber \\
& \hspace{20mm} + P[\textrm{node reg. due to recommendations}] \nonumber \\
 &=  f(q,u) \label{eq:u1_fixed_point}
\end{align}
where $f(q,u)$ is as given at the top of this page.

For any $q$, Equation~(\ref{eq:u1_fixed_point}) is a fixed point equation in $u$. However, due to the complex nature of the function $f(q,u)$, the existence of a $u$ that satisfies Equation~\eqref{eq:u1_fixed_point} is not obvious. {Using results from reliability theory, in Proposition~\ref{prop:fixed_point}, we prove the existence and uniqueness of the fixed point of Equation~\eqref{eq:u1_fixed_point}.} Before that, the following proposition establishes the nature of function $f(q,u)$ (w.r.t $u$).

\begin{prop} \label{prop:f_nature}
If the fraction of zealous nodes lies in the open interval $(0,1)$, then $ \forall \ q \in [0,1]$, $f(q,u)$ is a continuously differentiable, convex, monotonically increasing function of $u$.
\end{prop}
\begin{IEEEproof}
In Appendix~\ref{sec:prop_f_nature}.
\end{IEEEproof}

{Proposition~\ref{prop:f_nature} gives us some intuition about the existence and uniqueness of the fixed point.} However, we establish the same rigorously in the following proposition.
\begin{prop} \label{prop:fixed_point} 
For every $q \in [0,1]$, 
\begin{enumerate}[(i)]
\item If no node in the network is zealous, then $u=0$ is the only solution of Equation~\eqref{eq:u1_fixed_point}.
\item If every node in the network is zealous, then $u=1$ is the only solution of Equation~\eqref{eq:u1_fixed_point}.
\item If the fraction of zealous nodes lies in the interval $(0,1)$, then Equation~\eqref{eq:u1_fixed_point} has a unique fixed point in $(0,1)$.
\end{enumerate}
\end{prop}
\begin{IEEEproof}
In Appendix~\ref{sec:fixed_point}.
\end{IEEEproof}

For ease of presentation, in the remainder of the paper, we assume that the fraction of zealous nodes lies in the open interval $(0,1)$. For any $q \in [0,1]$, let $u_q$ denote the fixed point of Equation~\eqref{eq:u1_fixed_point}. {Since the fixed point is unique, we can obtain $u_q$ by iteratively solving Equation~\eqref{eq:u1_fixed_point}.} Our next proposition qualitatively describes the variation of the fixed point $u_q$ as function of $q$. 

\begin{prop} \label{prop:fp_increasing}
If $\alpha_2 > \alpha_1$, then $u_q$ is a strictly increasing continuous function of $q$, i.e., $\frac{\partial u_q}{\partial q} > 0$.
\end{prop}
\begin{IEEEproof}
In Appendix~\ref{sec:du_by_dq}.
\end{IEEEproof}

  Now, given a $q \in [0,1]$, let $s_k(q)$ be the {expected fraction} of degree $k$ nodes that have registered at the end of the campaign. The expression for $s_k(q)$ can be obtained as follows
  \begin{align}
&  s_k(q) = P[\textrm{node registers}| \textrm{degree } k ]  \nonumber \\
&=  P[\textrm{node is zealous} | \textrm{degree } k]  \nonumber \\
& \hspace{10mm} +  P[\textrm{node reg. due to recommendations} | \textrm{degree } k] \nonumber \\
&=  \sum \nolimits_{m \geq 1}  \hspace{-1mm} P[\textrm{node reg, node has threshold } m | \nonumber \\
& \hspace{20mm} \textrm{node has degree } k] +  p_{th}(0|k) \nonumber \\
&=  p_{th}(0|k) +  \sum \nolimits_{m \geq 1}  \hspace{-1mm} p_{th}(m|k) \cdot \nonumber \\
& \hspace{6mm} P[\textrm{node reg} | \textrm{node has degree } k,\textrm{ node has threshold } m]   \nonumber \\
&= g_k(q,u) |_{u=u_q} \label{eq:sk}  
   \end{align}
where $g_k(q, u)$ is as given at the top of this page and $u_q$ is the fixed point of Equation\eqref{eq:u1_fixed_point}.  Given a $q \in [0,1]$, let $s(q)$ be the {expected fraction} of nodes that have registered at the end of the campaign. $s(q)$ is also termed as the size of the epidemic. Then, using arguments similar to the ones used to derive Equation~\eqref{eq:s}, it can be shown that 
\begin{align}
s(q) &=\sum \nolimits_{k \geq 1} p(k) \cdot  g_k(q,u) \Big|_{u=u_q}  \hspace{-2mm} = \sum \nolimits_{k \geq 1} p(k) \cdot s_k(q) \label{eq:s}
\end{align}

{Once again using results from reliability theory, in Proposition~\ref{prop:g_increasing}, we establish the increasing and non-decreasing nature of functions $g(q,u)$ and $g_k(q,u)$ (with respect to $u$ and $q$). }

\begin{prop} \label{prop:g_increasing}
If $\alpha_2 > \alpha_1$, then 
\begin{enumerate}
	\item $\frac{\partial g(q,u)}{\partial q} > 0$ and $\frac{\partial g(q,u)}{\partial u} > 0$
	\item $\frac{\partial g_k(q,u)}{\partial q} \geq 0$ and $\frac{\partial g_k(q,u)}{\partial u} \geq 0$.
\end{enumerate}
\end{prop}
\begin{IEEEproof}
Similar to that of Propositions~\ref{prop:f_nature} and \ref{prop:increasing}.
\end{IEEEproof}

 
 \section{{Cost Minimization Under Cascade Size Constraint}} \label{sec:cm}
 
 Providing incentives is a costly affair. The campaigner may either be interested in minimizing the cost while guaranteeing that a given proportion of population registers, or in maximizing the registrations for a given cost budget. In this section, we look at the former problem.


\subsection{Cost of incentivization} \label{sec:cost_structure}
Incentives provided by the campaigner is usually a function of its degree because the number of potential recommendations  depend on the degree. Let $c_k$ be the cost of incentivizing a degree $k$ node. 
\emph{Incentivized nodes obtain incentives only if they registers}. Therefore, the  expected cost per incentivized degree $k$ node is given by $c_k \cdot s_k(q)$, where $s_k(q)$ (see Equation~\ref{eq:sk}) is the probability that an incentivized degree $k$ node registers. Let $\phi(k)$ denote the probability of incentivizing a degree $k$ node. Then, the average cost per node for incentive-policy $(\boldsymbol{\phi}=\{\phi(k), k \geq 1\})$ is given by $\sum_{k \geq 1} p(k) \phi(k) \times (\textrm{cost of node of deg. } k) = \sum_{k \geq 1}  p(k) \cdot c_k \cdot \phi(k)  \cdot s_k(q)$, where $q = \frac{1}{\overline{d}} \sum_{k\geq 1} k \cdot p(k) \cdot \phi(k)$.

\subsection{Problem formulation} \label{sec:problem_cm}
Minimizing the cost while providing guarantees on the number of expected registrations is appropriate for campaigns where the campaigners are mandated to achieve a given target. This problem can be mathematically formulated as follows   
\begin{align}
      & \hspace{10mm} \underset{\boldsymbol{0} \leq \boldsymbol{\phi} \leq \boldsymbol{1}} {\text{min}} \label{eq:min_problem}
       \ \ \ \  \sum \nolimits_{k \geq 1} c_k \cdot p(k) \cdot \phi(k)  \cdot s_k(q)  \\
      & \hspace{0mm} \textrm{Subject to:} \quad  s(q) \geq \gamma \textrm{ and } q = \frac{1}{\overline{d}} \sum \nolimits_{k\geq 1} k \cdot p(k) \cdot \phi(k) \nonumber 
   \end{align}
where $\gamma \in [0,\gamma_{max}]$ is the minimum expected fraction of registered individuals that must be achieved and $\overline{d}$ is the mean degree of the network. Here, $\gamma_{max}$ is the expected fraction of registered individuals obtained by incentivizing everyone. The expression for $s_k(q)$ (Equation~\eqref{eq:sk}) and $s(q)$ (Equation~\eqref{eq:s}) involve $u_q$, which is the solution to the fixed point Equation~\eqref{eq:u1_fixed_point}. Thus, it is not possible to apply traditional analytical techniques such as the \emph{Karush-Kuhn-Tucker (KKT) conditions} to solve the above problem. Furthermore, the problem may be non convex, and applying numerical techniques such as \emph{genetic algorithms}, or \emph{Markov Chain Monte Carlo} methods may not yield a globally optimal solution. 

\subsection{Solution approach} \label{sec:sa_cm}
In this section, we present transformations that allow us to efficiently compute the global optima of the cost minimization problem presented in the previous section. To help us with this, we state and prove the following proposition. {The intuition behind this proposition is that, as $\alpha_2 > \alpha_1$, increasing $q$ increases the proportion of type~$2$ nodes, which results in a higher $s(q)$.}
 
\begin{prop} \label{prop:s_increasing}
If $\alpha_2 > \alpha_1$, then function $s:[0,1] \to [\gamma_{min},\gamma_{max}]$ is a monotonically increasing bijection, where $\gamma_{max}=s(0)$ and $\gamma_{max}=s(1)$ are the expected fraction of registered individuals obtained by incentivizing nobody and everybody, respectively.
\end{prop}
\begin{IEEEproof}
We have 
\begin{align}
\frac{\partial s(q)}{\partial q} &= \frac{\partial \left(g(q,u) |_{u=u_q} \right)}{\partial q} \nonumber \\
& \overset{(a)}{=}  \frac{\partial g(q,u)}{\partial q} \Bigg|_{u=u_q} + \frac{\partial g(q,u)}{\partial u} \Bigg|_{u=u_q} \cdot \frac{\partial u_q}{\partial q} \nonumber \\
& =  \underset{\textrm{Proposition \ref{prop:g_increasing}}}{(>0)} + \underset{\textrm{Proposition \ref{prop:g_increasing}}}{(> 0)} \cdot \underset{\textrm{Proposition \ref{prop:fp_increasing}}}{(>0)}  \nonumber \\
& \hspace{-15mm} \Rightarrow \frac{\partial s(q)}{\partial q} > 0 \label{eq:sq_inc}
\end{align}
{where Equality~(a) follows from an application of the chain rule.} From Inequality~\eqref{eq:sq_inc}, we can see that $s(q)$ is a monotonically increasing function defined on the closed interval $[0,1]$. Hence, $s:[0,1]\to[\gamma_{min},\gamma_{max}]$ is a bijection.
\end{IEEEproof}

Proposition~\ref{prop:s_increasing} establishes that $s(q)$ is a monotonically increasing function of $q$ and is a bijection. Therefore,  we can replace the constraint $s(q) \geq \gamma$ with the constraint $q \geq q_{\gamma}$, where $q_{\gamma}$ is such that $s(q_\gamma) = \gamma$. {Since $s(q)$ is a monotonically increasing function of $q$, we can compute $q_{\gamma}$ by performing a line search over the closed interval $[0,1]$.} The following algorithm does this.
\begin{algorithm} \label{algo:q_gamm}
	\caption{Algorithm to compute $q_{\gamma}$}
	\begin{algorithmic}[1]
		\STATE Set $q_1 :=0$ and $q_2 :=1$ 
		\WHILE{$|s(q_1) - s(q_2)| > \epsilon$}
		\STATE $q_{mp} := (q_1 + q_2)/2$
		\STATE Solve equation $u = f(q_{mp},u)$ to obtain $u_{q_{mp}}$.
		\IF{$s(q_{mp}) < \gamma$}
		\STATE $q_1 := q_{mp}$
		\ELSE
		\STATE $q_2 := q_{mp}$
		\ENDIF
		\ENDWHILE
		\STATE \textbf{return} $q_{\gamma} := (q_1+q_2)/2$
	\end{algorithmic}
\end{algorithm}

Proposition~\ref{prop:s_increasing} does not tell us how the expected fraction of registered degree $k$ nodes ($s_k(q)$) responds to changes in $q$.  {While, the function $s_k(q)$ may not be a bijection for all values of $k$, we show that it is a \emph{non-decreasing} function of $q$, for all values of $k$.}

\begin{prop} \label{prop:sk_increasing}
If $\alpha_2 > \alpha_1$, then for all $k \geq 1$, $\frac{\partial s_k(q)}{\partial q} \geq 0$
\end{prop}
\begin{IEEEproof}
By first writing the total derivative in terms of the partial derivatives (see Proposition~\ref{prop:s_increasing}), and then by using the non-negativity of these partial derivatives (from Propositions \ref{prop:fp_increasing} and \ref{prop:g_increasing}).
\end{IEEEproof}
   
{Now, if $\gamma \leq \gamma_{min}=s(0)$, then the cascade constraint is met without incentivizing anyone. In such cases, the optimum incentive policy and the optimum cost is given by $\{\phi(k)=0, \forall k \geq 1\}$ and $0$, respectively. On the other hand, if $\gamma = \gamma_{max}=s(1)$, then the cascade constraint is met only if everyone is incentivized, i.e., $\{\phi(k)=1, \forall k \geq 1\}$, and the optimum cost is given by $\sum_{k \geq 1} c_k \cdot p(k) \cdot s_k(1)$.} For the cases when $\gamma \in (\gamma_{min}, \gamma_{max})$, {due to Propositions~\ref{prop:s_increasing}}, the cost minimization problem can be re-written as follows 
\begin{align}
& \hspace{0mm} P_1: \underset{\boldsymbol{0} \leq \boldsymbol{\phi} \leq \boldsymbol{1}} {\text{min}} \label{eq:min_problem_alternate}
\ \ \ \ \sum \nolimits_{k \geq 1} c_k \cdot p(k) \cdot \phi(k)  \cdot s_k(q)  \\
& \hspace{0mm} \textrm{Subject to:} \quad  q \geq q_{\gamma} \textrm{ and } q = \frac{1}{\overline{d}} \cdot \sum \nolimits_{k\geq 1} k \cdot p(k) \cdot \phi(k)\nonumber
\end{align}
where $q_{\gamma}$ is chosen such that $s(q_\gamma) = \gamma$. Since $\gamma \in (\gamma_{min}, \gamma_{max})$, we have $q_{\gamma} \in (0,1)$ (see Proposition~\ref{prop:s_increasing}).  {The next proposition establishes the existence of an optimal policy, of problem $P_1$, that satisfies all its the constraints, with equality.}
\begin{prop} \label{prop:eq_to_lp_stage_one}
	If $\alpha_2 > \alpha_1$, then problem $P_1$ has an optimal solution $\boldsymbol{\phi}^{opt} = \{\phi^{opt}_k , k \geq 1\}$ such that $\sum_{k\geq 1} k \cdot p(k) \cdot \phi^{opt}(k) = \overline{d} \cdot q_{\gamma}$
\end{prop}
\begin{IEEEproof}
	In Appendix~\ref{sec:eq_to_lp_stage_one}.
\end{IEEEproof}

{From Proposition~\ref{prop:eq_to_lp_stage_one}, we can see that problem $P_1$ has an optimal solution that satisfies all its constraints with equality. Therefore, to obtain a solution of problem $P_1$, we restrict our search to the set of policies that satisfy the constraints of problem $P_1$ with equality. Subsequently, a change of variable gives us the following linear program}
\begin{align}
      & \hspace{10mm} P_2:  \underset{\{\nu_k , k\geq 1\}} {\text{min}} \label{eq:min_problem_linear}
       \ \ \ \ \overline{d}  \cdot q_{\gamma}  \cdot \sum \nolimits_{k \geq 1} \nu_k  \cdot \mu_k(q_{\gamma})  \\
      & \hspace{0mm} \textrm{Subject to:} \quad \sum \nolimits_{k\geq 1} \nu_k = 1 \textrm{ and }	0 \leq \nu_k \leq \frac{k \cdot p(k)}{\overline{d} \cdot q_{\gamma}} \, \, \forall k \geq 1\nonumber 
   \end{align}
where $\mu_k(q_{\gamma}) = \frac{s_k(q_\gamma) \cdot  c_k}{k}$. {The following proposition establishes a crucial relation between the optimal solutions of problems $P_1$ and $P_2$. }
\begin{prop} \label{prop:eq_to_lp_stage_two}
Let $\{\nu^{opt}_k , k \geq 1\}$ be an optimal solution of problem $P_2$. If $\alpha_2 > \alpha_1$, then we can obtain an optimal solution of problem $P_1$ as follows
$$\phi^{opt}(k) = \begin{cases}
0 & \textrm{if } p(k) = 0 \\
\frac{\nu^{opt}_k \cdot \overline{d} \cdot q_{\gamma}}{k \cdot p(k)} & \textrm{otherwise}
\end{cases}  $$
\end{prop}
\begin{IEEEproof}
In Appendix~\ref{sec:eq_to_lp_stage_two}.
\end{IEEEproof}

{Therefore, to obtain an optimal solution of problem $P_1$, we just need to solve $P_2$ and apply Proposition~\ref{prop:eq_to_lp_stage_two}. Problem $P_2$ is a linear program and can be solved by any well-known LP-solver in polynomial time ($O((k_{max})^{3.5})$ time, where $k_{max}$ is the maximum node degree in the network)\cite{Karmarkar1984}. However, we exploit the rich structure of this problem and solve it in linearithmic time $O(k_{max} \cdot log(k_{\max}))$}

Let $\mathcal{S} = \{1, 2, \cdots, k_{max}\}$. Here, $k_{max}$ is the maximum degree of the network. Consider a permutation $\sigma:\mathcal{S} \to \mathcal{S}$ such that if $k_1 < k_2$, then $\mu_{\sigma(k_1)}(q_{\gamma}) \leq \mu_{\sigma(k_2)}(q_{\gamma})$, i.e., $\mu_{\sigma(1)}(q_{\gamma}) \leq \mu_{\sigma(2)}(q_{\gamma}) \leq \cdots \leq \mu_{\sigma(k_{max})}(q_{\gamma})$. The permutation $\sigma$ can be obtained in $O(k_{max} \cdot log(k_{\max}))$ time by sorting the set $\{\mu_k(q_{\gamma}), k \geq 1\}$ in ascending order. Using this order, we can obtain an optimal solution $\boldsymbol{\nu}^a$ of problem $P_2$ in $\Theta(k_{max})$ time by following the steps in Algorithm~\ref{algo:compute_nu}. 
\begin{algorithm}[h] 
\caption{Algorithm to compute optimal solution of $P_2$ \label{algo:compute_nu}}
\begin{algorithmic}[1]
\STATE $res := 1$ and $\boldsymbol{\nu}^a := [0, 0, \cdots , 0]$ --- a $1 \times k_{max}$ vector
\FOR{$j \in [1, 2, \cdots, k_{max}]$}
\IF{$res > 0$}
\STATE $\nu^a_{\sigma(j)} = \min \left\{ res, \frac{\sigma(j) \cdot p(\sigma(j))}{\overline{d} \cdot q_{\gamma}} \right\}$
\STATE $res := res - \nu^a_{\sigma(j)}$
\ENDIF
\ENDFOR
\STATE \textbf{return} $\boldsymbol{\nu}^a$
\end{algorithmic}
\end{algorithm}

In Algorithm~\ref{algo:compute_nu}, $\nu^a_k$ denotes the $k^{th}$ component of vector $\boldsymbol{\nu}^a$.  {We illustrate the principle behind Algorithm~\ref{algo:compute_nu} using a jar filling analogy. Variable $\nu^a_{\sigma(j)}$ can be interpreted as a jar with capacity   $\frac{\sigma(j) \cdot p(\sigma(j))}{\overline{d} \cdot q_{\gamma}}$.  The for loop in the algorithm is equivalent to filling these jars with water from another} {jar of unit capacity, in an ascending order, as given by the permutation $\sigma$.} From steps $1$, and $4-5$ of Algorithm~\ref{algo:compute_nu}, it is easy to see that $\boldsymbol{\nu}^a$ is a feasible solution of problem $P_2$. The optimality of $\boldsymbol{\nu}^a$ with respect to problem $P_2$ is established by the following proposition.

\begin{prop} \label{prop:opt_stage_two}
The vector $\boldsymbol{\nu}^a$ returned by Algorithm~\ref{algo:compute_nu} is an optimal solution of problem $P_2$.
\end{prop}
\begin{IEEEproof}
In Appendix~\ref{sec:opt_nu_a}.
\end{IEEEproof}

\section{{Cascade Size Maximization Under Budget Constraint}} \label{sec:csm}

In this section, we consider the challenge faced by a resource-constrained campaigner. The campaigner desires to maximize the campaign size, but is constrained by limited resources. Let $\overline{c}$ denote the limit on the expected incentivization cost. Then, we can formulate the cost-constrained cascade maximization problem as follows
\begin{align*}
      & \hspace{15mm} P_3: \underset{\boldsymbol{0} \leq \boldsymbol{\phi} \leq \boldsymbol{1}} {\text{max}} 
       \ \ \ \ s(q)  \\
      & \hspace{0mm} \textrm{Subject to:} \ \  \sum \nolimits_{k \geq 1} c_k  \cdot p(k)  \phi(k)  s_k(q) \leq \overline{c} \nonumber \\
     &	\hspace{5mm}  \textrm{ and } q = \frac{1}{\overline{d}} \sum \nolimits_{k\geq 1} k \cdot p(k) \cdot \phi(k)
   \end{align*}

\subsection{Solution approach} \label{sec:sa_crm}
Let $s^{opt}$ be the optimum value of problem $P_3$. From Proposition~\ref{prop:s_increasing}, we know that there exists a unique $q^{opt} \in (0,1)$ such that $s(q^{opt}) = s^{opt}$. {From Proposition~\ref{prop:s_increasing}, we know that $\frac{\partial s(q)}{\partial q}> 0$, i.e., $s(q)$ is a monotonically increasing bijection. Therefore, it can be argued that $q^{opt}$ can be obtained by solving the following optimization problem} 
\begin{align*}
      & \hspace{15mm} P_4: \underset{\boldsymbol{0} \leq \boldsymbol{\phi} \leq \boldsymbol{1}} {\text{max}} 
      \ \ \ \ q  	\\
            & \hspace{0mm} \textrm{Subject to:}  \sum \nolimits_{k \geq 1} c_k  \cdot p(k)  \phi(k)  s_k(q) \leq \overline{c}   \nonumber      \\
             &	\hspace{5mm} \textrm{ and } q = \frac{1}{\overline{d}} \sum \nolimits_{k\geq 1} k \cdot p(k) \cdot \phi(k)
\end{align*}

Let us define $\mu_k(q) = \frac{s_k(q) \cdot  c_k}{k}$ and $\nu_k(q) = \frac{k \cdot p(k) \cdot \phi(k)}{\overline{d} \cdot q}$. Then, problem $P_4$ can be re-written as follows
\begin{align*}
 & \hspace{30mm} P_5: \underset{\boldsymbol{\nu}, q \in (0,1)} {\text{max}}   \ \ \ \ q  \\
  & \hspace{15mm} \textrm{Subject to:} \quad  \overline{d}  \cdot q  \cdot \sum \nolimits_{k \geq 1} \nu_k(q)  \cdot \mu_k(q) \leq \overline{c} \nonumber     \\
  &	\hspace{0mm}	  \sum \nolimits_{k\geq 1} \nu_k(q) = 1 \quad \textrm{and} \quad 0 \leq \nu_k(q) \leq \frac{k \cdot p(k)}{\overline{d} \cdot q} \quad \forall \ k \geq 1\nonumber 
\end{align*}

\noindent
Now, given a $q \in(0,1)$, consider the following criteria 
\begin{enumerate}[(1)]
\item $\overline{d}  \cdot q  \cdot \sum_{k \geq 1} \nu_k(q)  \cdot \mu_k(q) \leq \overline{c} $
\item $ \sum_{k\geq 1} \nu_k(q) = 1$
\item $0 \leq \nu_k(q) \leq \frac{k  \cdot p(k)}{\overline{d} \cdot q} \quad \forall k \geq 1$
\end{enumerate}

If we can find a vector $\boldsymbol{\nu}$ that satisfies the above criteria, then the 2-tuple $(\boldsymbol{\nu},q)$ is a feasible solution of problem $P_5$. Let $\boldsymbol{\nu}^a(q)$ be the vector output by Algorithm~\ref{algo:compute_nu} when $q_{\gamma}$ is set to $q$. From Proposition~\ref{prop:opt_stage_two}, we know that this vector minimizes the LHS of criterion~(1) subject to criteria (2) and (3). Therefore, if we have $\overline{d}  \cdot q  \cdot \sum_{k \geq 1} \nu^a_k(q)  \cdot \mu_k(q) > \overline{c} $, then for the given value of $q$, no vector $\boldsymbol{\nu}$ can satisfy criterion~(1). Now, let $\boldsymbol{\nu}^a(q+ \Delta q)$ be the vector output by Algorithm~\ref{algo:compute_nu} when $q_{\gamma}$ is set to $q+\Delta q$. Then, for any $\Delta q > 0$, if $\overline{d}  \cdot q  \cdot \sum_{k \geq 1} \nu^a_k(q)  \cdot \mu_k(q) > \overline{c}$, then we have
\begin{align}
\overline{c} & < \overline{d}  \cdot q   \sum_{k \geq 1} \nu^a_k(q)  \cdot \mu_k(q) \overset{(a)}{\leq} \overline{d}  \cdot q  \cdot \sum_{k \geq 1} \nu^a_k(q+\Delta q)  \cdot \mu_k(q) \nonumber \\
& \overset{(b)}{\leq} \overline{d}  \cdot (q + \Delta q)    \cdot \sum \nolimits_{k \geq 1} \nu^a_k(q+\Delta q)  \cdot \mu_k(q+\Delta q) \label{eq:less}
\end{align}
{where Inequality~$(a)$ follows because the vector $\boldsymbol{\nu}^a(q)$ is an optimal solution and $\boldsymbol{\nu}^a(q+\Delta q)$ is just a feasible point at $q$, and  Inequality~$(b)$ follows because $\frac{\partial \mu_k(q)}{\partial q} = \frac{c_k}{k} \cdot \frac{\partial s_k(q)}{\partial q} \geq 0$ (see Proposition~\ref{prop:sk_increasing})}. From Inequality~\ref{eq:less} we can conclude that if there does not exist a vector $\boldsymbol{\nu}$ that  satisfies criteria (1)-(3) for some $q \in (0,1)$, then no vector satisfies these criteria for any $q^{'} \in (q,1)$. {Therefore, the optimal value $q^{opt}$ is given by the largest value of $q$ in $ [0,1]$ such that vector $\boldsymbol{\nu}^a(q)$ satisfies criteria~(1)}, and can be computed using the following algorithm


\begin{algorithm} 
\caption{Algorithm to compute optimum of $P_3$ \label{algo:q_opt}}
\begin{algorithmic}[1]
\STATE Set $q_1 :=0$ and $q_2 :=1$ 
\WHILE{$|s(q_1) - s(q_2)| > \epsilon$}
\STATE $q_{mp} := (q_1 + q_2)/2$
\STATE Set $q_{\gamma} = q_{mp}$ in Algorithm~\ref{algo:compute_nu} and obtain  $\boldsymbol{\nu}^a(q_{mp})$
\IF{$\overline{d}  \cdot q_{mp} \cdot \sum_{k \geq 1} \nu^a_k(q_{mp})  \cdot \mu_k(q_{mp}) > \overline{c}$}
\STATE $q_2 := q_{mp}$
\ELSE
\STATE $q_1 := q_{mp}$
\ENDIF
\ENDWHILE
\STATE Return $q^{opt} := (q_1+q_2)/2$
\end{algorithmic}
\end{algorithm}

Algorithm~\ref{algo:q_opt} does a line search to find the largest values of $q \in [0,1]$ such that the tuple $(\boldsymbol{\nu}^a(q),q)$ meets criteria (1)-(3). 

\section{Simulation and Numerical Evaluations} \label{sec:num_eval}
\begin{figure}[t]
	\centering
	\includegraphics[scale=0.25,angle=-90]{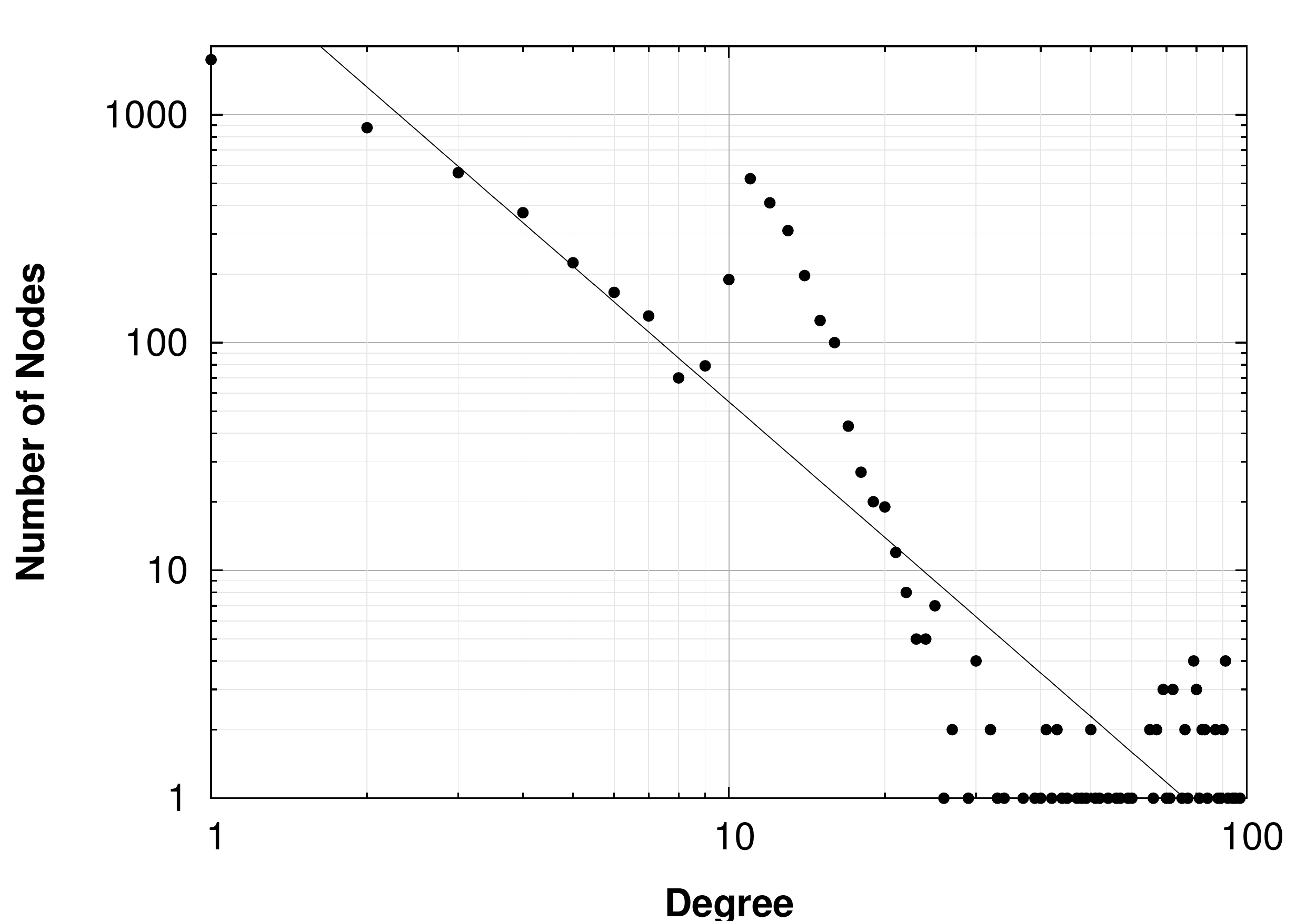}
	\caption{Degree distribution of network A (Gnutella network); the thick black line denotes the best fit of the power law to the degree distribution. \label{fig:gnu_deg_dist}}
	\label{fig:degreedistribution1}
	\includegraphics[scale=0.25,angle=-90]{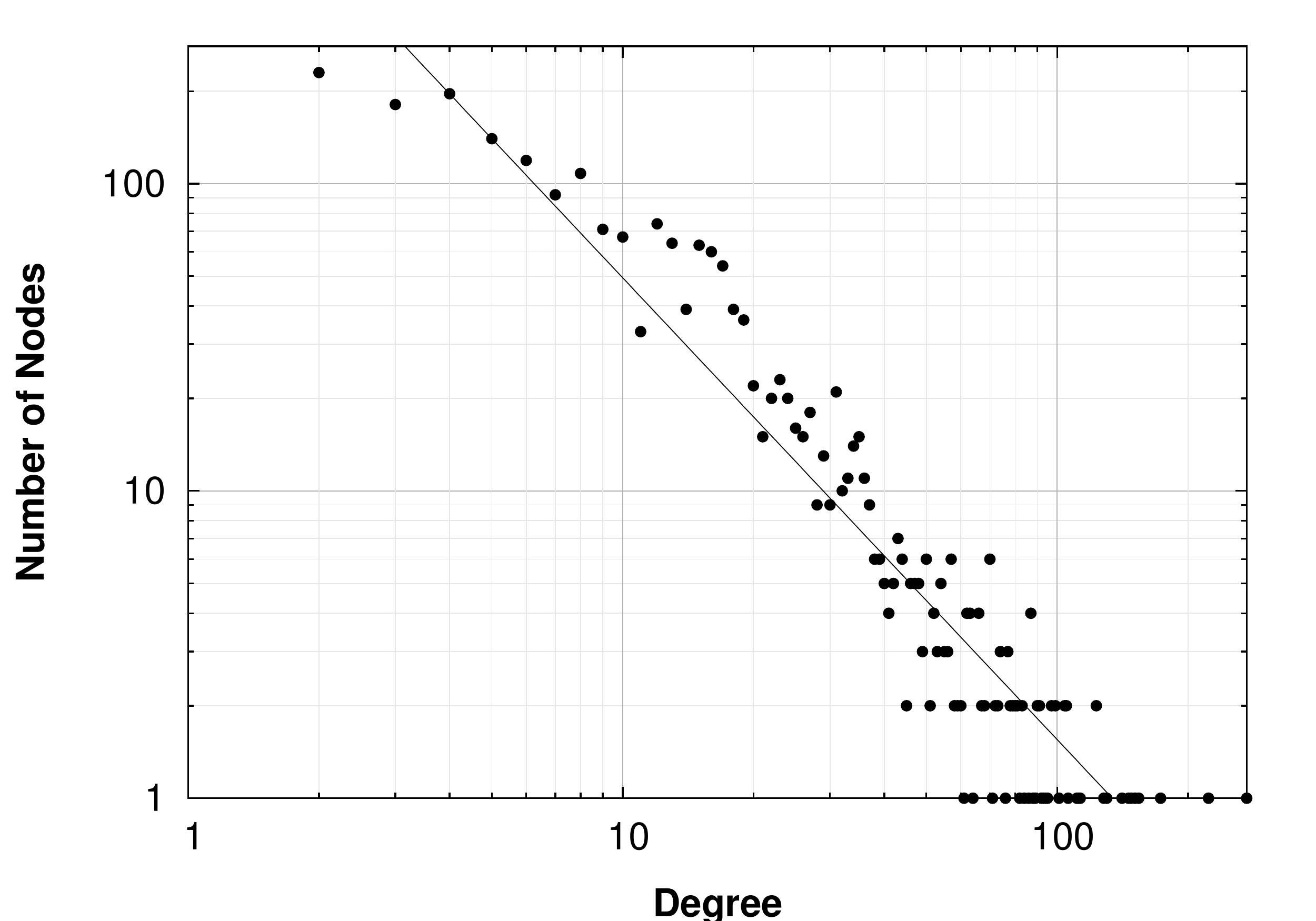}
	\caption{Degree distribution of network B (Hamsterster network); the thick black line denotes the best fit of the power law to the degree distribution. \label{fig:ham_deg_dist}}
	\label{fig:degreedistribution2}
\end{figure}

We formulated and solved the optimization problems assuming an uncorrelated, locally tree-like and connected network. However, real world social networks may contain triads, loops, multiple connected components, and need not be uncorrelated. Despite such differences, we show that, in most cases, our analytical results closely match the simulations on real world social networks. We evaluate our analytical results on two very different networks: Gnutella  \cite{snapnets} --- a p2p file sharing network, and Hamsterster \cite{konect} --- a social network for people with pet hamsters. Table~ \ref{table:networks} presents a few statistical measures of these networks. The degree distributions of these networks are displayed in Figures ~\ref{fig:degreedistribution1} and ~\ref{fig:degreedistribution2}.  All simulations are averaged over $10000$ runs.  To enhance legibility, error bars are suppressed in all the plots.

\begin{table}[h]
\caption{Simple Parameters of the two real-world networks used for simulations.}
\centering
\begin{tabular}{|c|c|c|}
\hline  & Network A &  Network B\\ 
\hline Source & p2p-Gnutella08 \cite{snapnets} & Hamsterster \cite{konect} \\
\hline Network category & Peer-to-peer & Social \\ 
\hline Nodes & 6301 & 2426 \\
\hline Edges & 20777 & 16631 \\
\hline Maximum degree & 273 & 97\\
\hline Average degree & 6.59 & 13.71\\
\hline Number of triangles & 2383 & 53265  \\
\hline Clustering coefficient &	0.01 & 0.51 \\
\hline Connected components & 2 & 148 \\
\hline 
\end{tabular} 
\label{table:networks}
\end{table}

\begin{figure}[t]
	\centering
	\includegraphics[scale=0.28,angle=-90]{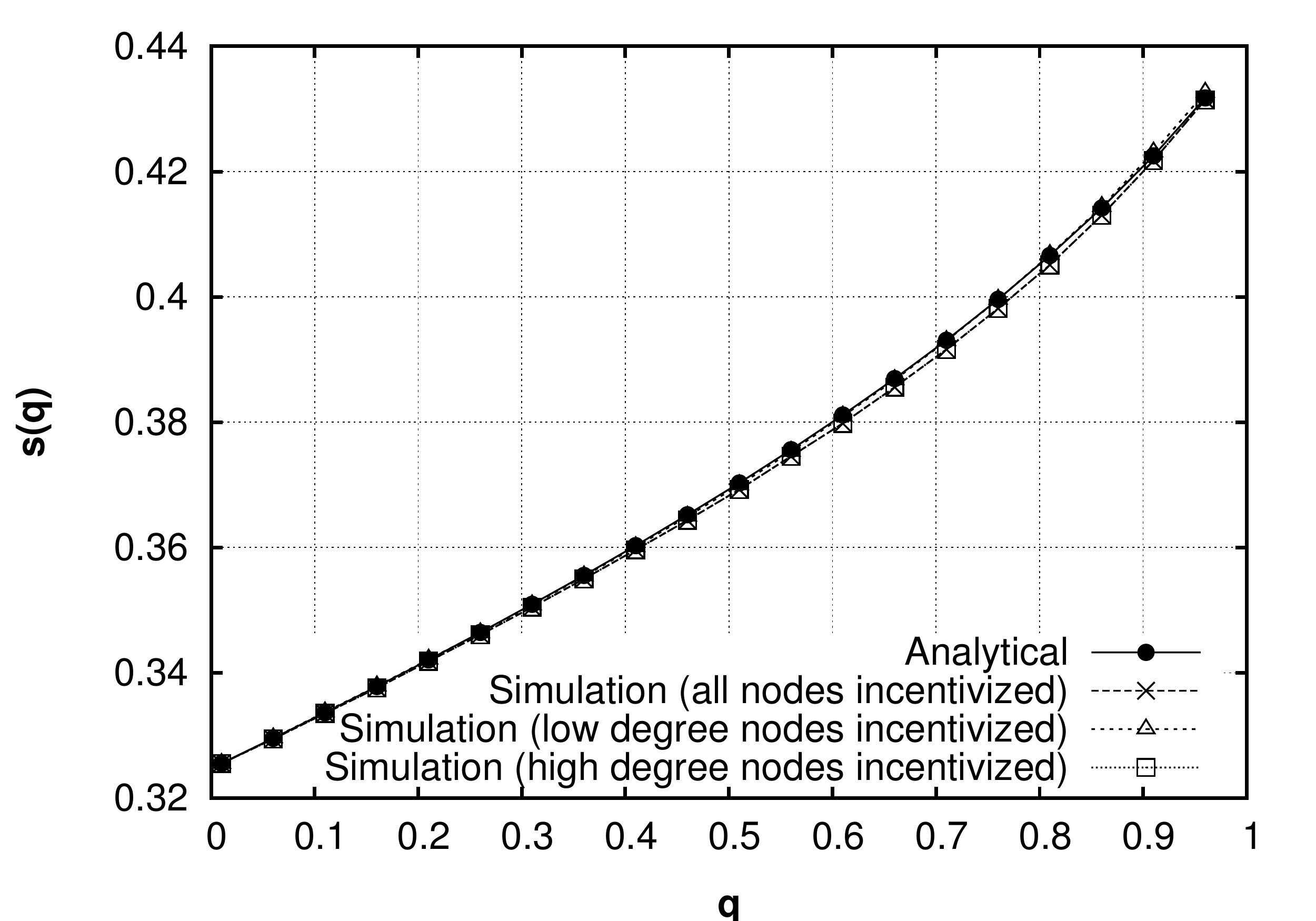}
	\caption{Analytical and simulated values of $s(q)$ on network A (Gnutella network). \label{fig:gnu_s_plot}}
	\label{fig:SimGnutella}
	\includegraphics[scale=0.28,angle=-90]{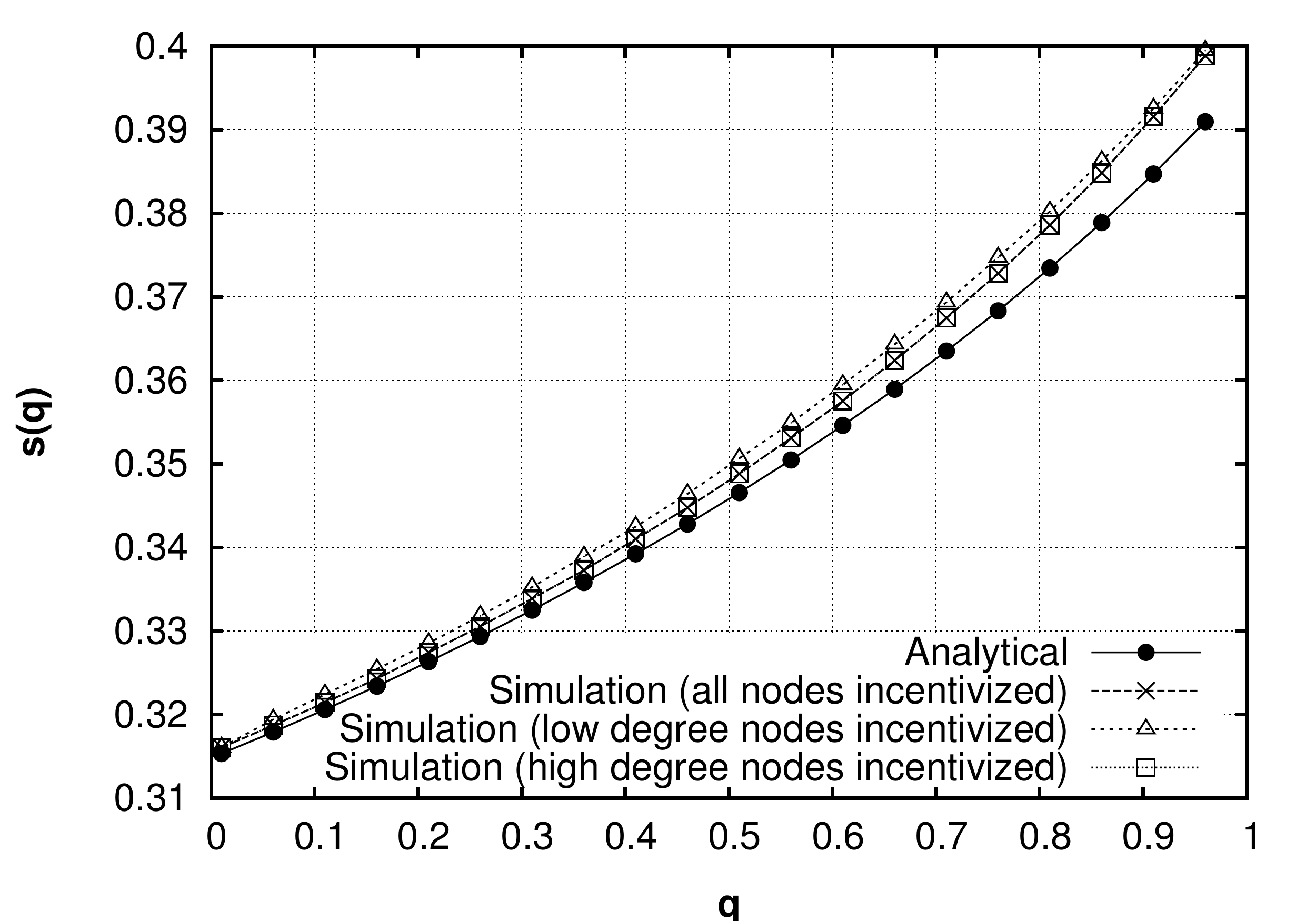}
	\caption{Analytical and simulated values of $s(q)$ on network B (Hamsterster network). \label{fig:ham_s_plot}}
	\label{fig:SimHamsterster}
	\centering
	\includegraphics[scale=0.345,angle=-90]{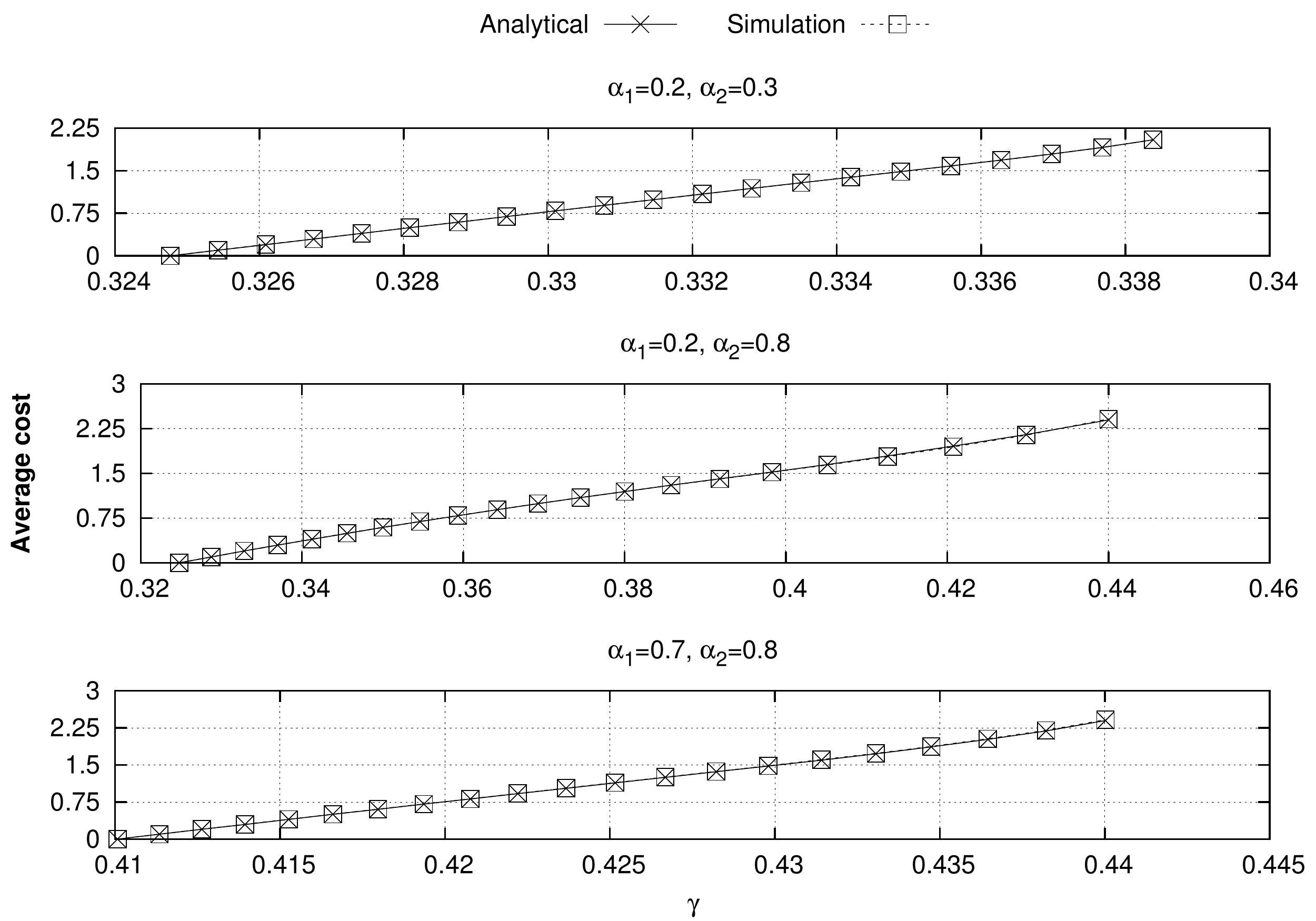}
	\caption{Analytical and simulated values of average cost vs. $\gamma$ for network A (Gnutella network). \label{fig:gnu_gamma_plot}}
	\label{fig:sim1gnutella}
\end{figure}

We first evaluate the correctness of the analytical calculation of the campaign size, i.e., $s(q)$.  The simulation plots were generated by considering a linear threshold model where nodes register if at least $50 \%$ of their neighbours are active. We assume that $30 \%$ of the nodes in the network were zealous.
 For a different values of $q$, we studied three schemes:  $\boldsymbol{\phi}^{all}$ --- incentivizing all nodes equally with probability $q$,  $\boldsymbol{\phi}^{high}$ --- incentivizing nodes starting from the highest degree till degree $k'$ such that $q=\frac{1}{\overline{d}} \sum_{k \geq k'} p(k) \phi^{high}(k)$,  and $\boldsymbol{\phi}^{low}$ --- incentivizing nodes starting from the lowest degree till degree $k'$ such that $q=\frac{1}{\overline{d}} \sum_{1 \leq k \leq k'} p(k) \phi^{high}(k)$. The analytical results were obtained by extracting the degree distribution from the networks. 
 
 The number of triangles in the p2p Gnutella network is very small, which is also reflected in the clustering coefficient. Therefore, the Gnutella network is very similar to a locally tree-like network. As consequence of this, for the Gnutella network, the simulation and analytical results are in excellent agreement with each other (see Fig.~\ref{fig:gnu_s_plot}). On the other hand, one can observe a deviation of the simulation results from the analytical on Hamsterster for large values of $q$ (see Fig.~\ref{fig:ham_s_plot}). This behaviour may in part be due to the presence of significant number of triangles and loops in the network.

\begin{figure}
\includegraphics[scale=0.35,angle=-90]{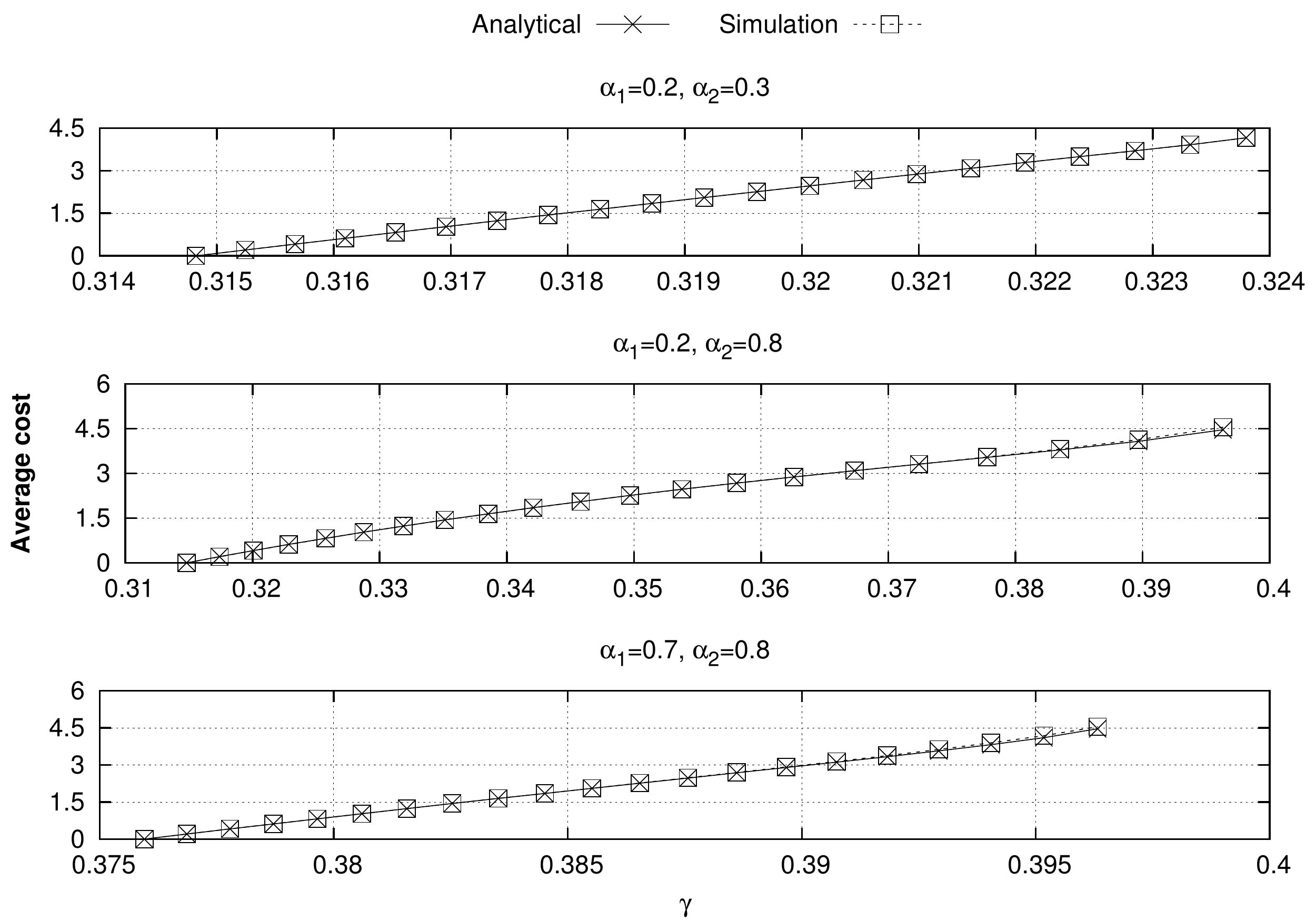}
\caption{Analytical and simulated values of average cost vs. $\gamma$ for network B (Hamsterster network). \label{fig:ham_gamma_plot}}
\label{fig:sim1hamsterster}
\centering
\includegraphics[scale=0.35,angle=-90]{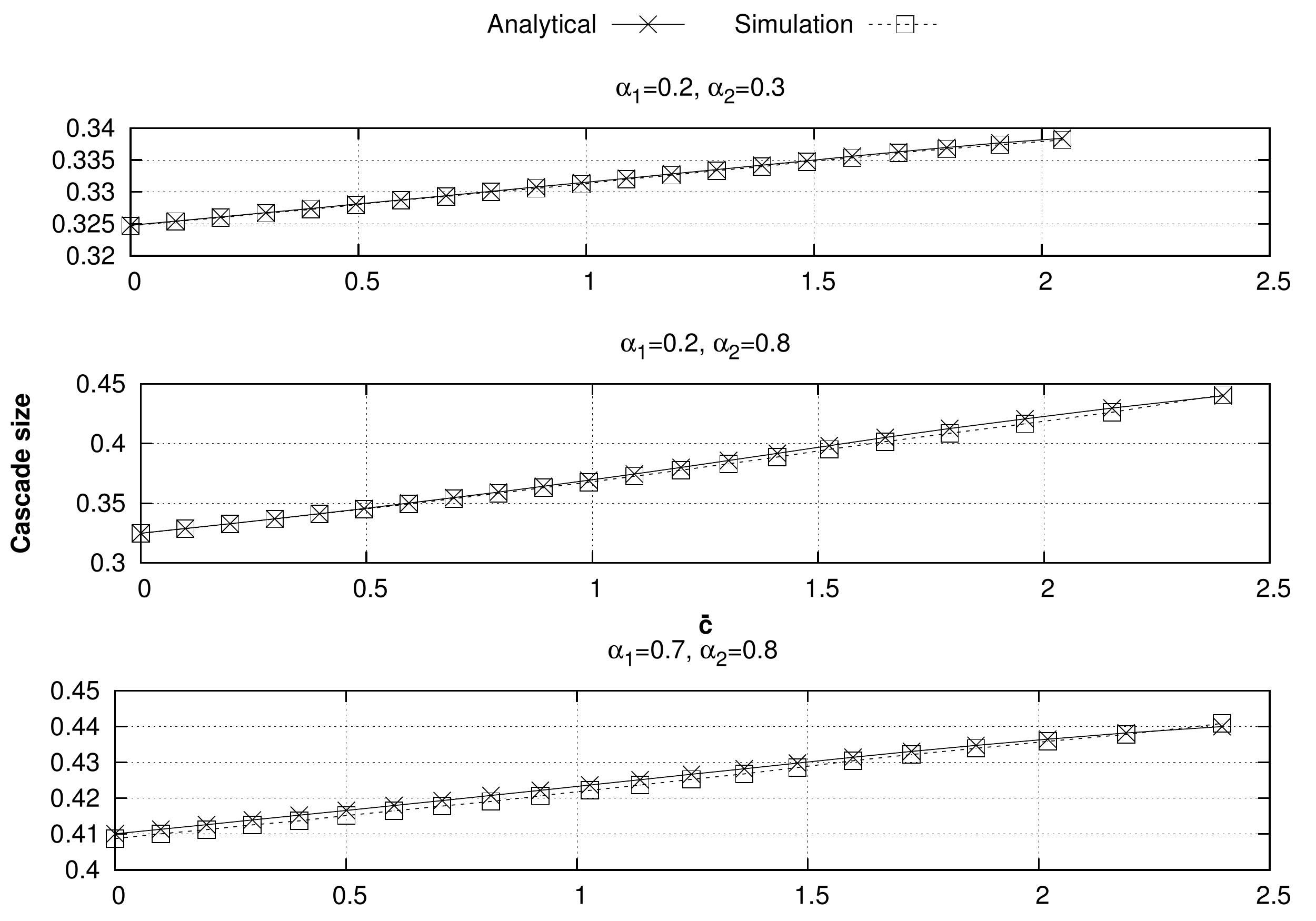}
\caption{Analytical and simulated values of cascade size vs. $\overline{c}$ for network A (Gnutella network). \label{fig:gnu_c_plot}}
\label{fig:sim2gnutella}
\end{figure}

\begin{figure}[t]
\includegraphics[scale=0.35,angle=-90]{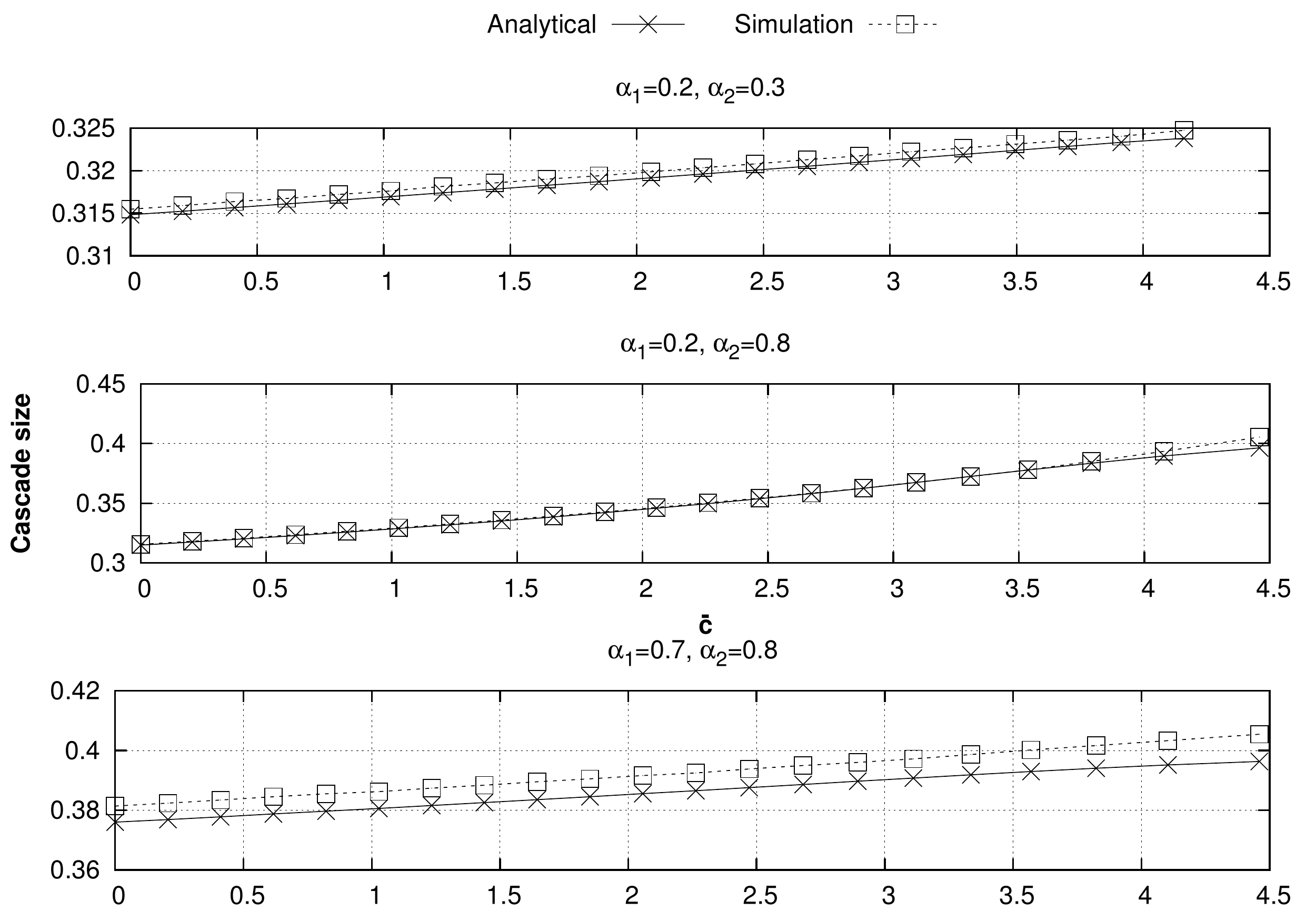}
\caption{Analytical and simulated values of cascade size vs. $\overline{c}$ for network B (Hamsterster network). \label{fig:ham_c_plot}}
\label{fig:sim2hamsterster}
\end{figure}

Next, 
we consider a linear incentive structure, where an incentivized node obtains a unit reward per neighbour, i.e., $c_k = k, \ \forall \ k \geq 1$. For the cost minimization problem, for a given $\gamma$, using the degree distributions, we analytically compute the solution $\boldsymbol \phi$ and the average cost. We then simulate the linear threshold process (nodes register if at least $50 \%$ of their neighbours are active) on the respective real world networks using the analytical solution $\boldsymbol{\phi}$, and compare the average cost obtained from analysis and simulations (see Fig.~\ref{fig:sim1gnutella} and Fig.~\ref{fig:sim1hamsterster}). Similarly, for the campaign size maximization problem we analytically compute $\boldsymbol{\phi}$ and size $s$ for a given cost budget $\overline{c}$. The analytical solution is then used in the simulation of cascade maximization problem (see Fig.~\ref{fig:sim2gnutella} and Fig.~\ref{fig:sim2hamsterster}).

As observed in Fig. ~\ref{fig:sim1gnutella} and ~\ref{fig:sim1hamsterster}, the analytical average cost is in excellent agreement with the simulated one for different values of  $\alpha_1, \alpha_2$ and $\gamma$, on both  the networks. Similarly Fig. ~\ref{fig:sim2gnutella} and ~\ref{fig:sim2hamsterster} show an excellent match between simulation and analytical computation of campaign size in almost all the plots. For large values of $\alpha_1$ and $\alpha_2$, the simulated campaign size in the Hamsterster network is larger than the analytically computed campaign size. This may in part be due to the large number of triangles in the network. This suggests that in social networks containing large number of triangles and loops, the analytically computed size represents a lower bound.   

\section{Conclusion} \label{sec:conclusion}
In this paper, we have studied the problem of campaigning  in social networks by offering incentives for referrals. We used ideas from percolation theory to compute the campaign size, which was then used to formulate two optimization problems.  These problems were not amenable to traditional solutions since they involved a fixed point equation whose solution was analytically intractable. We used results from reliability theory to establish some key properties of the fixed point that enabled us to solve these problems with simple algorithms having linearithmic time complexity. Although we assumed an uncorrelated and locally tree-like network in the analysis, through extensive simulations on real world social networks, we showed that our analytical results are applicable in real world networks. 


\appendices
\section{Some Results from Reliability Theory} \label{sec:tso}


In this paper, we prove several propositions using the theory of \emph{stochastic order.}. Let $X$ and $Y$ be two random variables. Then, $X$ is said to be smaller than $Y$ in the \emph{usual stochastic order} (denoted by $X \leq_{st} Y $) \cite{moshe} if and only if
$$ P[ X > x ] \leq  P[ Y > x ] \quad \forall x \in (-\infty, \infty)$$

Next, we present two theorem from \cite{moshe} without proof, and a lemma. We will use these theorems and lemma in several of our proofs.

\begin{thm} \label{thm:prop1_thm1}
$X \leq_{st} Y$ if and only if $\mathbb{E}[\psi(X)] \leq \mathbb{E}[\psi(Y)]$ holds for all non-decreasing functions $\psi$ for which the expectation exists.
\end{thm}
\begin{IEEEproof}
Refer Chapter~1 of \cite{moshe}.
\end{IEEEproof}

\begin{thm} \label{thm:prop1_thm2}
The usual stochastic order is closed under convolutions, i.e., If $X \leq_{st} Y$ and $X^{'} \leq_{st} Y^{'}$, then $X^{'} + X \leq_{st} Y^{'} + Y$.
\end{thm}
\begin{IEEEproof}
Refer Chapter~1 of \cite{moshe}.
\end{IEEEproof}

\begin{lem} \label{lemma:prop1_lemma1}
\begin{enumerate}[(i)]
\item Let $I(p_1)$ and $I(p_2)$ be two Bernoulli random variable with parameters $p_1$ and $p_2$, then $I(p_1) \leq_{st} I(p_2)$ if and only if $p_1 \leq p_2$. 
\item Let $Z(n, p_1)$ and $Z(n, p_2)$ be two binomial random variable with parameters $(n,p_1)$ and $(n,p_2)$, then for any $n \in \mathbb{N}$, $Z(n, p_1) \leq_{st} Z(n, p_2)$ if $p_1 \leq p_2$. 
\end{enumerate}
\end{lem}
\begin{IEEEproof}
\textit{Part (i):} By comparing the complementary cumulative distribution function of the two random variable, and by applying the definition of \emph{usual stochastic order}.

\textit{Part (ii):} A binomial random variable with parameter $(n,p)$ is the sum of $n$ i.i.d. Bernoulli random variable with parameter $p$. Now, the result follows from an application of Part~(i), and due to the fact that stochastic order is closed under convolution (Theorem~\ref{thm:prop1_thm2}).
\end{IEEEproof}


We also use some results from the theory of \emph{stochastic convexity} in our proofs. For the sake of completeness, we reproduce some definitions and theorems from \cite{moshe}. Let $\{X(\theta), \theta \in \Theta\}$ be a parametrized collection of random variables. We say
\begin{enumerate}[(a)]
\item $\{X(\theta), \theta \in \Theta\}$ is \emph{stochastically increasing (SI)} if $\mathbb{E}[\psi(X(\theta))]$ is non-decreasing in $\theta$ for all non-decreasing functions $\psi$.
\item $\{X(\theta), \theta \in \Theta\}$ is \emph{stochastically increasing and convex (SICX)} if $\{X(\theta), \theta \in \Theta\} \in SI$ and $\mathbb{E}[\psi(X(\theta))]$ is non-decreasing and convex in $\theta$ for all non-decreasing convex functions $\psi$.
\item $\{X(\theta), \theta \in \Theta\}$ is \emph{stochastically increasing and linear (SIL)} if $\{X(\theta), \theta \in \Theta\} \in SICX$ and $\mathbb{E}[\psi(X(\theta))]$ is non-decreasing and concave in $\theta$ for all non-decreasing concave functions $\psi$.
\end{enumerate}
\noindent
Note that, by definition, we have $\{X(\theta), \theta \in \Theta\} \in SIL \Rightarrow \{X(\theta), \theta \in \Theta\} \in SICX$. 

\begin{lem} \label{lemma:binomial_sicx}
Let $X(n,p)$ be a binomial random variable with parameters $(n,p)$. Then, $\{X(n,p), p \in (0,1) \} \in SICX$. 
\end{lem}
\begin{IEEEproof}
By combining example Example~8.B.3 and Theorem~8.B.9 of \cite{moshe}, it can shown that $\{X(n,p), p \in (0,1)\} \in SIL$. Then, due to the fact that $SIL \Rightarrow SICX$, the lemma follows.
\end{IEEEproof}

\begin{thm} \label{thm:closure_sicx}
Suppose $\{X(\theta), \theta \in \Theta\}$ and $\{Y(\theta), \theta \in \Theta\}$ are two collection of random variables such that $X(\theta)$ and $Y(\theta)$ are independent for each $\theta$. If $\{X(\theta), \theta \in \Theta\} \in SICX$ and $\{Y(\theta), \theta \in \Theta\} \in SICX$, then $\{X(\theta)+Y(\theta), \theta \in \Theta\} \in SICX$.
\end{thm}
\begin{IEEEproof}
Refer the proof of Theorem~8.A.15 in \cite{moshe}.
\end{IEEEproof}

\begin{thm} \label{thm:sicx_and_ccdf}
Suppose that for each $\theta \in \Theta$, the support of $X(\theta)$ is in $\mathbb{N}$. Then, $\{X(\theta), \theta \in \Theta\} \in SICX$ \emph{if and only if} $\{X(\theta), \theta \in \Theta\} \in SI$ and $P[X(\mathbf{\theta}) > m]$ is non-decreasing and convex in $\theta$ for all $m \in \mathbb{N}$.
\end{thm}
\begin{IEEEproof}
Refer the proof of Theorem~8.A.10 in \cite{moshe}.
\end{IEEEproof}

\section{Existence and Uniqueness of the  Fixed Point} \label{sec:fixed_point_proof}
In this section, we show that for any $q \in [0,1]$, the fixed point Equation~\ref{eq:u1_fixed_point} has a unique solution. The proof in this section is split into two parts. In the first part, we show that function $f(q,u)$ is monotonically increasing and convex in $u$. In the section part, we use the result of the first part to establish the existence and uniqueness of the fixed point.

	\subsection{Proof of Proposition~\ref{prop:f_nature}} \label{sec:prop_f_nature}
The differentiability of $f(q,u)$ follows from its definition. In this section, we show that $f(q,u)$ is monotonically increasing and convex in $u \in (0,1)$. To show this, we use the theory of stochastic convexity. 

Let $\{X(k_2, \alpha_2 u), u \in (0,1) \}$ and $\{Y(k-k_2,\alpha_1u), u \in (0,1) \}$ be two collection of independent binomial random variables with parameters $(k_2,\alpha_2u)$ and $(k-k_2,\alpha_1u)$, respectively. 

Then, from Lemma~\ref{lemma:binomial_sicx}, we have $\{X(k_2, \alpha_2 u), u \in (0,1) \} \in SICX$ and $\{Y(k-k_2,\alpha_1u), u \in (0,1) \} \in SICX$. Since the random variables are independent, an application of Theorem~\ref{thm:closure_sicx} gives us $\{X(k_2, \alpha_2 u) + Y(k-k_2,\alpha_1u), u \in (0,1) \} \in SICX$. Now, by applying Theorem~\ref{thm:sicx_and_ccdf} we can conclude that
\begin{enumerate}[(1)]
\item $\{X(k_2, \alpha_2 u) + Y(k-k_2,\alpha_1u), u \in (0,1) \} \in SI$.
\item $P[X(k_2, \alpha_2 u) + Y(k-k_2,\alpha_1u) > m]$ is a \emph{non-decreasing convex} function of $u \in (0,1)$ for all $m \in \mathbb{N}$.
\end{enumerate}

Let $h(u,k,k_2,m) = P[X(k_2, \alpha_2 u) + Y(k-k_2,\alpha_1u) > m]$. We note that function $h(u,k,k_2,m)$ is non-decreasing and convex in $u$. Since the network is connected and the fraction of zealous nodes lies in the open interval $(0,1)$, there exists positive integers $k_0,m_0$ such that $m_{0} \in (0, k_0]$ and $p_{ex}(k_0) p_{th}(m_0 | k_0+1) > 0$. Therefore, we have
\begin{align*}
& \frac{\partial f(q, u)}{\partial u} = \sum_{k \geq 1} p_{ex}(k) \sum_{m \geq 1} p_{th}(m|k+1) \cdot \sum^{k}_{k_2=0} \hat{p}(k_2|k) \cdot \\
& \hspace{10mm} \frac{\partial h(u,k,k_2,m)}{\partial u} \cdot (1 - \mathbb{I}_{\{k=k_0,m=m_0,k_2=k\}}) \\
& \hspace{5mm} + p_{ex}(k_0) p_{th}(m_0|k_0+1) \frac{\partial P[X(k_0, u\alpha_2) \geq m_0]}{\partial u} \overset{(a)}{>} 0
\end{align*}
where Inequality~(a) follows because $h(u,k,k_2,m)$ and $P[X(k_0, u\alpha_2) \geq m_0], m_0 \in (0, k_0]$ are non-decreasing and monotonically increasing functions of $u$, respectively. Similarly, we have 
\begin{align*}
&\frac{ \partial^2 f(q, u)}{\partial u^2} = \sum_{k \geq 1} p_{ex}(k) \sum_{m \geq 0} p_{th}(m|k+1) \sum^{k}_{k_2=0} \hat{p}(k_2|k) \cdot \\
&\hspace{30mm} \frac{\partial^2 h(u,k,k_2,m)}{\partial u^2} \overset{(b)}{\geq} 0
\end{align*}
where Inequality~(a) follows because $h(u,k,k_2,m)$ is convex in $u$. \hfill \IEEEQED

\subsection{Proof of Proposition~\ref{prop:fixed_point}} \label{sec:fixed_point}
We recollect that $u$ denotes the probability of finding a registered node by following an arbitrary link of the network. Nodes with zero threshold do not need any recommendations, and are registered from the start of the campaign. Thus, if every node has threshold value zero, then the entire network is registered and we have $u=1$. On the other hand, nodes with non-zero threshold have to be recommended by active nodes to register. Therefore, if all nodes in the network have non-zero threshold, then there are no node from which the campaign can start. Hence, in such scenarios, we have $u=0$. 

Now, let us consider the case when the fraction of zealous nodes lies in the interval $(0,1)$. Let us define $f^0(u) = f(q,u) - u$. It is easy to see that $u_0$ is a fixed point of Equation~\ref{eq:u1_fixed_point} \emph{if and only if} $f^{0}(u_0)=0$. By differentiating $f^0(u)$ with respect to $u$ twice, we get $\frac{\partial^2 f^0(u)}{\partial u^2} = \frac{\partial^2 f(q,u)}{\partial u^2} \geq 0$, i.e., $f^0$ is a convex function of $u$. We also have
\begin{align*}
f^0(0) &= f(q,0) = \sum \nolimits_{k \geq 0} p_{ex}(k) \cdot p_{th}(0|k+1) \overset{(a)}{>} 0 \\
f^0(1) &= f(q,1) - 1 \overset{(b)}{<} 0
\end{align*} 
where Inequality~(a) follows because the network is connected and the fraction of zealous nodes lies in the open interval $(0,1)$, and Inequality~(b) follows because the $f(q,1)$ is a convex combination of non-negative terms less or equal to $1$ (some terms are strictly less that $1$). Since $f^0(0) > 0$ and $f^0(1) < 0$, due to the continuity of function $f^{0}$, there exist a point $u_0 \in (0,1)$ such that $f^{0}(u_0)=0$ i.e., $f(q,u_0) = u_0$. 

We prove the uniqueness of the fixed point by contradiction. Now, if $u_0, u_1 \in (0,1)$ are two fixed points of Equation~\eqref{eq:u1_fixed_point}, then we should have $f^{0}(u_0) = f^{0}(u_1) = 0$. Without loss of generality, let us assume that $0 < u_0 < u_1 < 1$. Now, choose positive real numbers $u_2 \in (u_1,1)$ and $\lambda \in (0,1)$ such that $f^0(u_2)<0$ and $u_1 = \lambda u_0 + (1-\lambda)u_2$. Then, we have 
\begin{align*}
&f^0(\lambda u_0 + (1-\lambda)u_2) = f^0(u_1) = 0 \quad \textrm{and}\\
&\lambda f^0(u_0) + (1-\lambda) f^0(u_2) = (1-\lambda) f^0(u_2) < 0 \\
& \Longrightarrow f^0(\lambda u_0 + (1-\lambda)u_2) \overset{(c)}{>} \lambda f^0(u_0) + (1-\lambda) f^0(u_2)
\end{align*}

Inequality~(c) contradicts the convex nature of function $f^{0}$. Thus, in turn, establishes the uniqueness of the fixed point. \hfill \IEEEQED

\section{Increasing Nature of the Fixed Point} \label{sec:du_by_dq}

First we introduce some notation and present some preliminary observations. We will use these observations in the proof of Proposition~\ref{prop:fp_increasing}. Let 
$$h_{m,k}(q) = \sum^{k}_{k_2=0} {k \choose k_2}  q^{k_2}  (1-q)^{k - k_2}   P[X_{k_2} + Y_{k-k_2} \geq m ]$$
where random variables $X_{k_2}$ and $Y_{k-k_2}$ are independent and have a binomial distribution with parameters $(k_2,\alpha_2 u)$ and $(k-k_2,\alpha_1 u)$, respectively. The first step in the proof of Proposition~\ref{prop:fp_increasing} is to show that for any $m \in \mathbb{Z}^{+}$, $P[X_{k_2} + Y_{k-k_2} \geq m ]$ is a non-decreasing function of $k_2 \in [0,k]$. Let 
$$\varphi_{m,k}(k_2) = P[X_{k_2} + Y_{k-k_2} \geq m] \quad \forall k_2 \in [0,k]$$

We note that if $u \in (0,1]$ then $\varphi_{m,k}(k_2)  = 0$ only if $m > k$, i.e., if the number of active neighbours is greater than the node degree.
\begin{lem} \label{lemma:prop1_lemma2}
For any $m,k \in \mathbb{Z}^{+}$, $\varphi_{m,k}(k_2)$ is non-decreasing function of $k_2 \in [0, k]$. 
\end{lem}
\begin{IEEEproof}
Consider two integers $k_2,k^{'}_2 \in [0,k]$. Without loss of generality, let $k^{'}_2 > k_2$ and define the following random variables
$$W_{k_2} = X_{k_2} + Y_{k-k_2} \quad \textrm{and} \quad  W_{k^{'}_2} = X_{k^{'}_2} + Y_{k-k^{'}_2}$$
Then, we have
\begin{align*}
W_{k_2} &=_{st} X_{k_2} + Y_{k-k_2} \\ 
&=_{st} \sum \nolimits^{k_2}_{i=1} I_i(\alpha_2 u) + \sum \nolimits^{k-k_2}_{i=1} I_i(\alpha_1 u)
\end{align*}
where $I_i(p)$ are \emph{independent Bernoulli random variable} with parameter $p$ and $=_{st}$ implies equality in distribution \cite{moshe}. Therefore, for any $k^{'}_2 > k_2$ and $k_2,k^{'}_2 \in [0,k]$, we have
\begin{align*}
& W_{k_2} =_{st} \sum\nolimits^{k_2}_{i=1} I_i(\alpha_2 u) + \sum\nolimits^{(k-k^{'}_2) + (k^{'}_2-k_2)}_{i=1} I_i(\alpha_1 u) \\
&=_{st} \sum\nolimits^{k_2}_{i=1} I_i(\alpha_2 u) + \sum\nolimits^{k-k^{'}_2}_{i=1} I_i(\alpha_1 u)  +  \sum\nolimits^{k^{'}_2-k_2}_{i=1} I_i(\alpha_1 u) \\
&\overset{(a)}{\leq}_{st} \sum\nolimits^{k_2}_{i=1} I_i(\alpha_2 u) + \sum\nolimits^{k-k^{'}_2}_{i=1} I_i(\alpha_1 u) +  \sum\nolimits^{k^{'}_2-k_2}_{i=1} I_i(\alpha_2 u) \\
& =_{st} \sum\nolimits^{k^{'}_2}_{i=1} I_i(\alpha_2 u) + \sum\nolimits^{k-k^{'}_2}_{i=1} I_i(\alpha_1 u) =_{st} W_{k^{'}_2}
\end{align*}
where Inequality $(a)$ follows because $\alpha_1 < \alpha_2 \Rightarrow \alpha_1 u \leq  \alpha_2 u$, and due to Lemma~\ref{lemma:prop1_lemma1}. Hence, we have established that for $k_2,k^{'}_2 \in [0,k]$ and $k^{'}_2 > k_2$, we have $W_{k_2}\leq_{st} W_{k^{'}_2}$. Thus, $\forall m \in \mathbb{Z}^{+}$ and $k^{'}_2 > k_2$, we have
\begin{align*}
\varphi_{m,k}(k_2) = P[X_{k_2} + Y_{k-k_2} \geq m] &= P[W_{k_2} \geq m] \\
& \hspace{-45mm} \overset{(b)}{\leq}  P[W_{k^{'}_2} \geq m]  = P[X_{k^{'}_2} + Y_{k-k^{'}_2} \geq m] = \varphi_{m,k}(k^{'}_2)
\end{align*}
where Inequality (b) follows because $W_{k_2}\leq_{st} W_{k^{'}_2}$. Thus, we have shown that $\varphi_{m,k}(k_2)$ is a non-decreasing function of $k_2 \in [0, k]$.
\end{IEEEproof} 
Now, let us define
$$\psi_{m,k}(k_2) = \frac{k \cdot \varphi_{m,k}(k_2)}{(k-k_2)} \quad \forall k_2 \in [0,k-1]$$

Due to Lemma~\ref{lemma:prop1_lemma2}, we can conclude that for any $m,k \in \mathbb{Z}^{+}$, $\psi_{m,k}(k_2)$ is also a \emph{non-decreasing function} of $k_2 \in [0, k-1]$. Next, establish that the function $f(q,u)$ is monotonically increasing in $q$. 

\begin{prop} \label{prop:increasing}
If $\alpha_2 > \alpha_1$, then $\frac{\partial f(q,u)}{\partial q} > 0$.
\end{prop}
\begin{IEEEproof}
\begin{align*}
& h_{m,k}(q) = \sum^{k}_{k_2=0} {k \choose k_2} q^{k_2} (1-q)^{k - k_2} P[X_{k_2} + Y_{k-k_2} \geq m ] \\
&= \sum\nolimits^{k-1}_{k_2=0} {k \choose k_2} q^{k_2} (1-q)^{k - k_2} P[X_{k_2} + Y_{k-k_2} \geq m ] \\
& \hspace{40mm} + q^m \cdot P[X_{k} \geq m ] \\
&= (1-q) \sum\nolimits^{k-1}_{k_2=0} {k-1 \choose k_2} q^{k_2} (1-q)^{k - k_2} \cdot \psi_{m,k}(k_2) \\
& \hspace{40mm} + q^m \cdot P[X_{k} \geq m ] \\
&= (1-q) \cdot \mathbb{E}[\psi_{m,k}(Z(k-1,q))] + q^m \cdot P[X_{k} \geq m ] 
\end{align*}
where $Z(k-1,q)$ is a binomial random variable with parameters $(k-1,q)$. 

Now, consider two real numbers $q_1,q_2 \in [0,1]$ such that $q_1 < q_2$. Then, we have
\begin{align}
& h_{m,k}(q_1) =  (1-q_1) \mathbb{E}[\psi_{m,k}(Z(k-1,q_1))] \nonumber \\
& \hspace{50mm} + (q_1)^m P[X_{k} \geq m ] \nonumber \\
& \overset{(a)}{\leq}  (1-q_1) \mathbb{E}[\psi_{m,k}(Z(k-1,q_1))] + (q_2)^m P[X_{k} \geq m ]  \nonumber \\
& \overset{(b)}{\leq}  (1-q_1) \mathbb{E}[\psi_{m,k}(Z(k-1,q_2))] + (q_2)^m P[X_{k} \geq m ]  \nonumber \\
& \overset{}{=}  (1-q_2) \mathbb{E}[\psi_{m,k}(Z(k-1,q_2))] + (q_2)^m P[X_{k} \geq m ]  \nonumber \\
& \hspace{20mm} + (q_2-q_1) \mathbb{E}[\psi_{m,k}(Z(k-1,q_2))] \nonumber \\
& \overset{}{=}  h_{m,k}(q_2) + (q_2-q_1) \mathbb{E}[\psi_{m,k}(Z(k-1,q_2))] \label{eq:h_m_one}
\end{align}
where \begin{enumerate}[(a)]
\item follows because $q_1 < q_2$.
\item since $q_1 < q_2$, we have  $Z(k-1,q_1) \leq_{st} Z(k-1,q_2)$ (Lemma~\ref{lemma:prop1_lemma1}). Therefore, due to the non-decreasing nature of function $\psi_{m,k}$, we have $\mathbb{E}[\psi_{m,k}(Z(k-1,q_1))] \leq \mathbb{E}[\psi_{m,k}(Z(k-1,q_2))]$ (Theorem~\ref{thm:prop1_thm1}). 
\end{enumerate}
Also, from the definition, we also have
\begin{align}
h_{m,k}(q_1) & = \mathbb{E}[\varphi_{m,k}(Z(k,q_1))] \nonumber \\
&\overset{(c)}{\leq} \mathbb{E}[\varphi_{m,k}(Z(k,q_2))] = h_{m,k}(q_2) \label{eq:h_m_two}
\end{align}
where Inequality $(c)$ follows because of Lemma~\ref{lemma:prop1_lemma1}, Lemma~\ref{lemma:prop1_lemma2} and Theorem~\ref{thm:prop1_thm1}.

From Equations~\eqref{eq:h_m_one} and \eqref{eq:h_m_two}, we can see that for any $q_1,q_2 \in [0,1]$ such that $q_1 < q_2$, we have $h_{m,k}(q_1)= h_{m,k}(q_2)$ \emph{if and only if} $\mathbb{E}[\psi_{m,k}(Z(k-1,q_2))]=0$. From the definition of function $\psi_{m,k}$, we can see that $\mathbb{E}[\psi_{m,k}(Z(k-1,q_2))] = 0$ \emph{if and only if} $m > k$. Thus, if $m \in [0,k]$, then we can conclude that $h_{m,k}(q_1) < h_{m,k}(q_2)$. 

Since the network is connected and the fraction of zealous nodes lies in the open interval $(0,1)$, there exists positive integers $k_0,m_0$ such that $m_{0} \in (0, k_0]$ and $p_{ex}(k_0) p_{th}(m_0 | k_0+1) > 0$. Therefore,  for $q_1,q_2 \in [0,1]$ such that $q_1 < q_2$, we have $h_{m_0,k_0}(q_1) < h_{m_0,k_0}(q_2) $. Let $z_0 = \sum\nolimits_{k \geq 1} p_{ex}(k)   p_{th}(0|k+1) $ --- fraction of zealous nodes. Thus, we have
\begin{align*}
&f(q_1,u) = z_0 + \sum_{k \geq 1} \sum_{m \geq 1} p_{ex}(k)   p_{th}(m|k+1) h_{m,k}(q_1) \\
& = \sum_{k \geq 1} \sum_{m \geq 1} p_{ex}(k)   p_{th}(m|k+1) h_{m,k}(q_1) (1-\mathbb{I}_{\left\{\underset{k=k_0}{m=m_0}\right\}}) \\
& \hspace{20mm} +  p_{ex}(k_0)   p_{th}(m_0|k_0+1) \cdot h_{m_0,k_0}(q_1) +  z_0\\
& < \sum_{k \geq 1} \sum_{m \geq 1} p_{ex}(k)   p_{th}(m|k+1) h_{m,k}(q_1) (1-\mathbb{I}_{\left\{\underset{k=k_0}{m=m_0}\right\}}) \\
& \hspace{20mm} +  p_{ex}(k_0)   p_{th}(m_0|k_0+1) \cdot h_{m_0,k_0}(q_2) + z_0 \\
& \leq \sum_{k \geq 1} \sum_{m \geq 1} p_{ex}(k)   p_{th}(m|k+1) h_{m,k}(q_2) (1-\mathbb{I}_{\left\{\underset{k=k_0}{m=m_0}\right\}}) \\
& \hspace{20mm} +  p_{ex}(k_0)   p_{th}(m_0|k_0+1) \cdot h_{m_0,k_0}(q_2) + z_0\\
& = f(q_2,u)
\end{align*}
i.e., for $q_1,q_2 \in [0,1]$ such that $q_1 < q_2$, we have $f(q_1,u) < f(q_2,u)$. 
\end{IEEEproof}

\subsection{Proof of Proposition ~\ref{prop:fp_increasing}}
The continuity of the fixed point follows from the fact that function $f(\cdot,u)$ is a continuous map for all $u \in [0,1]$. To prove the increasing nature of the fixed point, we use the \emph{implicit function theorem}. 

Let $f^{0}(q,u) = f(q,u) - u$. According to the implicit function theorem, when $f^0(q,u)=0$, we have
$$ \frac{du}{d q} = - \frac{\frac{\partial f^0(q,u)}{\partial q}}{\frac{\partial f^0(q,u)}{\partial u}} =  \frac{\frac{\partial f(q,u)}{\partial q}}{ \left(1 -  \frac{\partial f(q,u)}{\partial u} \right)} $$

From Proposition~\ref{prop:fixed_point}, we know that for every $q \in [0,1]$, there is a unique fixed point $u_q \in (0,1)$ that satisfies $f(q,u_q) = u_q$ i.e., $f^0(q,u_q)=0$. Therefore, we have 
$$ \frac{du_q}{d q} = \frac{\frac{\partial f(q,u_q)}{\partial q}}{ \left(1 -  \frac{\partial f(q,u)}{\partial u} \Big|_{u = u_q} \right)} $$

From Proposition~\ref{prop:increasing}, we know that $\frac{\partial f(q,u)}{\partial q} > 0 , \forall u \in (0,1)$. Since $u_q \in (0,1)$, we have $\frac{\partial f(q,u_q)}{\partial q} > 0$. Now, to prove Proposition ~\ref{prop:fp_increasing}, we just need to show that $\frac{\partial f(q,u)}{\partial u} \Big|_{u = u_q} < 1$. We show this by contradiction. We note that $f(q,1) \overset{(a)}{<} 1$, where Inequality~(a) follows because the $f(q,1)$ is a convex combination of non-negative terms less than or equal to $1$ (some  terms are strictly less that $1$). Since the function $f(q,u)$ is continuous and non-decreasing in $u$ (see Proposition~\ref{prop:f_nature}), there exists a point $u^{'}_q \in (u_q,1)$ such that $f(q,u^{'}_q) < u^{'}_q$. 

We note that the function $f(q,u)$ is convex in $u \in (0,1)$ (see Proposition~\ref{prop:f_nature}). Now, let us assume that $\frac{\partial f(q,u)}{\partial u} \Big|_{u = u_q} \geq 1$.  Then, from the definition of convex functions, we have
\begin{align*}
f(q,u^{'}_q) &\geq f(q,u_q) + (u^{'}_q - u_q) \cdot \frac{\partial f(q,u)}{\partial u} \Big|_{u = u_q} \\
&\overset{}{\geq} f(q,u_q) + (u^{'}_q - u_q) \\
&\overset{(b)}{=} u_q + (u^{'}_q - u_q) = u^{'}_q \\
& \hspace{-20mm} \Rightarrow f(q,u^{'}_q) \overset{(c)}{\geq}  u^{'}_q
\end{align*}
where Inequality~(b) follows because we have $f(q,u_q) = u_q$. 

Inequality~(c) contradicts the fact that $f(q,u^{'}_q) < u^{'}_q$. Since this contradiction was due to the assumption that $\frac{\partial f(q,u)}{\partial u} \Big|_{u = u_q} \geq 1$, we should have $\frac{\partial f(q,u)}{\partial u} \Big|_{u = u_q} < 1$. This, in turn, establishes that $\frac{du_q}{d q} > 0$. \hfill \IEEEQED

\section{Equivalence to the Linear Program} \label{sec:eq_to_lp}
In this section, we prove the equivalence between problems $P_1$ and $P_2$ presented in Section~\ref{sec:cm}

\subsection{Proof of Proposition~\ref{prop:eq_to_lp_stage_one}} \label{sec:eq_to_lp_stage_one}

Let the value of the objective of problem $P_1$ for incentive-policy $\boldsymbol{\phi}$ be denoted as $c(\boldsymbol{\phi})$. Let $\boldsymbol{\phi}^{o} = \{\phi^{o}_k , k \geq 1\}$ be an optimal solution of problem $P_1$. If we have $\sum_{k\geq 1} k \cdot p(k) \cdot \phi^{o}(k) = \overline{d} \cdot q_{\gamma}$, then we can choose $\boldsymbol{\phi}^{opt} = \boldsymbol{\phi}^{o}$. 

So, let us assume that $\frac{1}{\overline{d}} \sum_{k\geq 1} k \cdot p(k) \cdot \phi^{o}(k) = q^o > q_{\gamma}$. Our goal to obtain a policy $\boldsymbol{\phi}^{opt}$ from $\boldsymbol{\phi}^{o}$ without increasing the cost. Let $\boldsymbol{\phi}^{opt} = \frac{q_{\gamma}}{q^o} \boldsymbol{\phi}^{o} $. Then, we have
\begin{align}
&\frac{1}{\overline{d}} \sum_{k\geq 1} k  p(k) \cdot \phi^{opt}(k) =  \frac{q_{\gamma}}{q^o \overline{d}} \sum_{k\geq 1} k \cdot p(k) \cdot \phi^{o}(k) = q_{\gamma} \label{eq:one} \\
& \hspace{20mm} \textrm{and} \quad  \boldsymbol{0} \leq \boldsymbol{\phi}^{opt} \leq \boldsymbol{\phi}^{o} \leq \boldsymbol{1} \label{eq:two}
\end{align}

From Equality~\eqref{eq:one} and Inequality~\eqref{eq:two} we can see that $\boldsymbol{\phi}^{opt}$ is a feasible solution of problem $P_1$. Further, we also have
\begin{align}
c(\boldsymbol{\phi}^{o}) &=  \sum \nolimits_{k \geq 1} c_k \cdot p(k) \cdot \phi^o(k)  \cdot s_k(q^o) \nonumber \\
& \overset{(a)}{\geq}  \sum \nolimits_{k \geq 1} c_k \cdot p(k) \cdot \phi^{opt}(k)  \cdot s_k(q^o) \nonumber \\
& \overset{(b)}{\geq}  \sum \nolimits_{k \geq 1} c_k \cdot p(k) \cdot \phi^{opt}(k)  \cdot s_k(q_{\gamma}) \nonumber \\
& = c(\boldsymbol{\phi}^{opt}) \overset{(c)}{\geq} c(\boldsymbol{\phi}^{o})  \Rightarrow c(\boldsymbol{\phi}^{o}) = c(\boldsymbol{\phi}^{opt}) \nonumber
\end{align}
where Inequality~(a) follows because $\boldsymbol{\phi}^{opt} \leq \boldsymbol{\phi}^{o}$, Inequality~(b) follows because $q^0 > q_{\gamma}$ and $s_k(q)$ is non-decreasing in $q$ (Proposition~\ref{prop:sk_increasing}), and Inequality~(c) follows because $\boldsymbol{\phi}^{o}$ is an optimal solution of problem $P_1$.

Since the cost for policies $\boldsymbol{\phi}^{o}$ and $\boldsymbol{\phi}^{opt}$ are equal, and $\boldsymbol{\phi}^{o}$ is an optimal policy, we can conclude that policy $\boldsymbol{\phi}^{opt}$ is also an optimal solution of problem $P_1$. \hfill \IEEEQED

\subsection{Proof of Proposition~\ref{prop:eq_to_lp_stage_two}} \label{sec:eq_to_lp_stage_two}
Let $\boldsymbol{\nu}^{opt} = \{\nu^{opt
}_k , k \geq 1\}$ be an optimal solution of problem $P_2$. Let us define
$$\phi^{f}(k) = \begin{cases}
0 & \textrm{if } p(k) = 0 \\
\frac{\nu^{opt}_k \cdot \overline{d} \cdot q_{\gamma}}{k \cdot p(k)} & \textrm{otherwise}
\end{cases}  $$
Then, we have
\begin{align}
\frac{1}{\overline{d}} \sum \nolimits_{k \geq 1} k \cdot p(k) \cdot \phi^{f}(k) =  q_{\gamma} \sum\nolimits_{k \geq 1} \nu^{opt}_k \overset{(a)}{=} q_{\gamma} \label{eq:feasible_phi_1} \\
\phi^{f}(k) = \begin{cases}
0 < 1 & \textrm{if } p(k) = 0 \\
\frac{\nu^{opt}_k \cdot \overline{d} \cdot q_{\gamma}}{k \cdot p(k)} \overset{(b)}{\leq} 1 & \textrm{otherwise}
\end{cases} \label{eq:feasible_phi_2} 
\end{align}
where Equality~(a) and Inequality~(b) follows because $\{\nu^{opt}_k , k \geq 1\}$ is a feasible solution of problem $P_2$. From Equality~\eqref{eq:feasible_phi_1} and Inequality~\eqref{eq:feasible_phi_2}, we can conclude that $\boldsymbol{\phi}^f$ is a feasible solution of problem $P_1$.

Now, let $\boldsymbol{\phi}^{opt} = \{\phi^{opt}_k , k \geq 1\}$ be an optimal solution of problem $P_1$ such that $\sum_{k\geq 1} k \cdot p(k) \cdot \phi^{opt}(k) = \overline{d} \cdot q_{\gamma}$. The existence of such an optimal policy is guaranteed by Proposition~\ref{prop:eq_to_lp_stage_one}. Let us define
$$\nu^{f}(k) = \frac{k \cdot p(k) \cdot \phi^{opt}_k  }{\overline{d} \cdot q_{\gamma}} $$
Then, we have $\sum_{k \geq 1} \nu^{f}(k) = \sum_{k \geq 1} \frac{k \cdot p(k) \cdot \phi^{opt}_k  }{\overline{d} \cdot q_{\gamma}} = 1$ and $0 \leq \nu_k =  \frac{k \cdot p(k) \cdot \phi^{opt}_k  }{\overline{d} \cdot q_{\gamma}} \leq \frac{k \cdot p(k)}{\overline{d} \cdot q_{\gamma}}$. Hence, $\boldsymbol{\nu}^f$ is a feasible solution of problem $P_2$. Further, we also have 
\begin{align*}
& \sum \nolimits_{k \geq 1} c_k   p(k)  \phi^{opt}(k)  s_k(q_{\gamma}) \leq \sum \nolimits_{k \geq 1} c_k   p(k)  \phi^{f}(k)  s_k(q_{\gamma})\\
& = \overline{d}  \cdot q_{\gamma}   \sum \nolimits_{k \geq 1} \nu^{opt}_k \cdot \mu_k(q_{\gamma})  \leq \overline{d}  \cdot q_{\gamma}   \sum \nolimits_{k \geq 1} \nu^{f}_k \cdot \mu_k(q_{\gamma}) \\
& =   \sum_{k \geq 1} c_k  p(k)  \phi^{opt}(k)  s_k(q_{\gamma}) \\
& \Rightarrow \sum_{k \geq 1} c_k  p(k)  \phi^{opt}(k)  s_k(q_{\gamma}) =  \sum_{k \geq 1} c_k  p(k)  \phi^{f}(k)  s_k(q_{\gamma}) \\
& \Rightarrow \boldsymbol{\phi}^f \textrm{ is an optimal solution of problem $P_1$.}
\end{align*}  \hfill \IEEEQED

\section{Optimality of Algorithm~\ref{algo:compute_nu}} \label{sec:opt_nu_a}
In this section, we prove that the vector $\boldsymbol{\nu}^a$ returned by Algorithm~\ref{algo:compute_nu} is an optimal point of problem $P_2$. However, as a first step, we prove the following lemma.
Let $\mathcal{S} = \{1, 2, \cdots, k_{max}\}$ and $k_{max}$ is the maximum degree in the network.
\begin{lem} \label{lem:opt_stage_one}
Let $\boldsymbol{\nu}$, a $1 \times k_{max}$ vector, be a feasible solution of problem $P_2$. Then, we have
$$\sum \nolimits^{j_0}_{j = 1} \nu^a_{\sigma(j)} \geq \sum \nolimits^{j_0}_{j = 1} \nu_{\sigma(j)} \quad \forall j_0 \in \mathcal{S}$$

\end{lem}
\begin{IEEEproof}
We prove this lemma by induction. For $j_0=1$, we have
\begin{align}
\nu^a_{\sigma(1)} = \min \left\{ 1, \frac{\sigma(1) \cdot p(\sigma(1))}{\overline{d} \cdot q_{\gamma}} \right\} \overset{(a)}{\geq} \nu_{\sigma(1)}
\end{align}
where Inequality~(a) follows because $\boldsymbol{\nu}$ is a feasible solution of problem $P_2$. Now, let us assume that $\sum^{j_1}_{j = 1} \nu^a_{\sigma(j)} \geq \sum^{j_1}_{j = 1} \nu_{\sigma(j)}$ for some $j_1 \in \mathcal{S}$. We need to show that $\sum^{j_1+1}_{j = 1} \nu^a_{\sigma(j)} \geq \sum^{j_1+1}_{j = 1} \nu_{\sigma(j)}$. 

If we have $\nu^a_{\sigma(j_1+1)}\geq \nu_{\sigma(j_1+1)}$, then the result is trivial. Let us therefore consider the case where $\nu^a_{\sigma(j_1+1)} < \nu_{\sigma(j_1+1)}$. This implies that
\begin{align*}
&\nu^a_{\sigma(j_1+1)} <  \nu_{\sigma(j_1+1)} \leq \frac{\sigma(j_1+1) \cdot p(\sigma(j_1+1))}{\overline{d} \cdot q_{\gamma}} \\
&\Rightarrow \nu^a_{\sigma(j_1+1)} \overset{(b)}{=} 1 - \sum \nolimits^{j_1}_{j=1} \nu^a_{\sigma(j)}
\end{align*}
where Equality~(b) follows because of steps~4 and 5 of Algorithm~\ref{algo:compute_nu}. Therefore, if $\nu^a_{\sigma(j_1+1)} < \nu_{\sigma(j_1+1)}$, then we have
\begin{align*}
\sum \nolimits^{j_1+1}_{j=1} \nu^a_{\sigma(j)} = 1 = \sum \nolimits_{j \in  \mathcal{S}} \nu_{\sigma(j)} \geq \sum \nolimits^{j_1+1}_{j = 1} \nu_{\sigma(j)}
\end{align*}
\end{IEEEproof}

\subsection{Proof of Proposition~\ref{prop:opt_stage_two}}
Let $\boldsymbol{\nu}$ be a feasible solution of problem $P_2$. Since $\boldsymbol{\nu}^a$ and $\boldsymbol{\nu}$ are feasible solutions of problem $P_2$, we have $\sum_{j \geq 1} \nu^a({\sigma(j)}) =  \sum_{j \geq 1} \nu({\sigma(j)}) = 1$. Let $X^a$ and $X$ two independent random variable taking values from the set $\mathcal{S}$ such that $P[X^a = j] = \nu^a_{\sigma(j)} \quad \textrm{and} \quad  P[X = j] = \nu_{\sigma(j)}$. Then, from Lemma~\ref{lem:opt_stage_one}, we have 
\begin{align}
& P[X^a \leq j_0] = \sum_{j=1}^{j_0} \nu^a_{\sigma(j)} \geq  \sum_{j=1}^{j_0} \nu_{\sigma(j)} = P[X \leq j_0] \label{eq:final}
\end{align}

From Equation~\ref{eq:final} and the definition of usual stochastic orders we can see that $X^a \leq_{st} X$. Therefore, we have
\begin{align}
& \sum \nolimits_{k \geq 1} \nu^a_k  \cdot  \mu_{k}(q_{\gamma}) = \sum \nolimits_{j \in \mathcal{S}} \nu^a_{\sigma(j)} \cdot  \mu_{\sigma(j)}(q_{\gamma}) \nonumber \\ 
&= \sum \nolimits_{j \in \mathcal{S}} P[X^a = j] \cdot  \mu_{\sigma(j)}(q_{\gamma}) = \mathbb{E}[\mu_{\sigma{(X^a)}}(q_{\gamma})] \nonumber \\
& \overset{(a)}{\leq} \mathbb{E}[\mu_{\sigma{(X)}}(q_{\gamma})] = \sum \nolimits_{j \in \mathcal{S}} P[X = j] \cdot  \mu_{\sigma(j)}(q_{\gamma}) \nonumber \\
&= \sum \nolimits_{j \in \mathcal{S}} \nu_{\sigma(j)} \cdot \mu_{\sigma(j)}(q_{\gamma}) = \sum \nolimits_{k \geq 1} \nu_k \cdot  \mu_{k}(q_{\gamma}) \label{eq:va}
\end{align}
where Inequality~(a) follows from an application of Theorem~\ref{thm:prop1_thm1} ($X^a \leq_{st} X$ and $\mu_{\sigma(j)}(q_{\gamma})$ is a non-decreasing function of $j$). 

Let $\boldsymbol{\nu}^{opt}$ be an optimal solution of $P_2$. Then,  we have the following inequalities 
\begin{align*}
& \sum_{k \geq 1} \nu^{opt}_k \cdot \mu_{k}(q_{\gamma}) \leq \sum_{k \geq 1} \nu^a_k \cdot \mu_{k}(q_{\gamma}) \overset{(b)}{\leq} \sum_{k \geq 1} \nu^{opt}_k \cdot \mu_{k}(q_{\gamma}) \\
& \Longrightarrow \sum_{k \geq 1} \nu^{opt}_k \cdot \mu_k (q_{\gamma}) = \sum_{k \geq 1} \nu^{a}_k \cdot  \mu_{k}(q_{\gamma}) \\
& \Longrightarrow \boldsymbol{\nu}^{a}  \textrm{ is an optimal solution of } P_2
\end{align*}
where Inequality~(b) follows by setting $\boldsymbol{\nu}$ to $\boldsymbol{\nu}^{opt}$ in Inequality~\eqref{eq:va}. \hfill \IEEEQED 

\bibliographystyle{IEEEtran}
\footnotesize
\bibliography{Information_Spread_LT}

\begin{thebibliography}{10}
\providecommand{\url}[1]{#1}
\csname url@samestyle\endcsname
\providecommand{\newblock}{\relax}
\providecommand{\bibinfo}[2]{#2}
\providecommand{\BIBentrySTDinterwordspacing}{\spaceskip=0pt\relax}
\providecommand{\BIBentryALTinterwordstretchfactor}{4}
\providecommand{\BIBentryALTinterwordspacing}{\spaceskip=\fontdimen2\font plus
\BIBentryALTinterwordstretchfactor\fontdimen3\font minus
  \fontdimen4\font\relax}
\providecommand{\BIBforeignlanguage}[2]{{%
\expandafter\ifx\csname l@#1\endcsname\relax
\typeout{** WARNING: IEEEtran.bst: No hyphenation pattern has been}%
\typeout{** loaded for the language `#1'. Using the pattern for}%
\typeout{** the default language instead.}%
\else
\language=\csname l@#1\endcsname
\fi
#2}}
\providecommand{\BIBdecl}{\relax}
\BIBdecl

\bibitem{Leskovec2009}
J.~Leskovec, L.~Backstrom, and J.~Kleinberg, ``Meme-tracking and the dynamics
  of the news cycle,'' in \emph{Proc. 15th ACM SIGKDD Int. Conf. Knowledge
  Discovery and Data Mining}, 2009, pp. 497--506.

\bibitem{Leskovec2006}
J.~Leskovec, A.~Singh, and J.~Kleinberg, ``Patterns of influence in a
  recommendation network,'' in \emph{Advances in Knowledge Discovery and Data
  Mining}.\hskip 1em plus 0.5em minus 0.4em\relax Springer, 2006, pp. 380--389.

\bibitem{Dinh2012}
T.~N. Dinh, D.~T. Nguyen, and M.~T. Thai, ``Cheap, easy, and massively
  effective viral marketing in social networks: truth or fiction?'' in
  \emph{Proc. 23rd ACM Conf. Hypertext and Social Media}, 2012, pp. 165--174.

\bibitem{Houston2010}
D.~Houston, ``Dropbox - startup lessons learned,'' \emph{Presentation Slides
  avaliable at
  http://www.slideshare.net/gueste94e4c/dropbox-startup-lessons-learned-3836587},
  2010.

\bibitem{Kempe2003}
D.~Kempe, J.~Kleinberg, and E.~Tardos, ``Maximizing the spread of influence
  through a social network,'' in \emph{Proc. 9th ACM SIGKDD Int. Conf.
  Knowledge Discovery and Data Mining}, 2003, pp. 137--146.

\bibitem{Chen2009}
W.~Chen, Y.~Wang, and S.~Yang, ``Efficient influence maximization in social
  networks,'' in \emph{Proc. 15th ACM SIGKDD Int. Conf. Knowledge Discovery and
  Data Mining}, 2009, pp. 199--208.

\bibitem{Chen2010}
W.~Chen, Y.~Yuan, and L.~Zhang, ``Scalable influence maximization in social
  networks under the linear threshold model,'' in \emph{Proc. 10th IEEE Int.
  Conf. Data Mining (ICDM)}, 2010, pp. 88--97.

\bibitem{Hartline2008}
J.~Hartline, V.~Mirrokni, and M.~Sundararajan, ``Optimal marketing strategies
  over social networks,'' in \emph{Proc. 17th Int. Conf. World Wide Web}, 2008,
  pp. 189--198.

\bibitem{Arthur2009}
D.~Arthur, R.~Motwani, A.~Sharma, and Y.~Xu, ``Pricing strategies for viral
  marketing on social networks,'' in \emph{Internet and Network
  Economics}.\hskip 1em plus 0.5em minus 0.4em\relax Springer, 2009, pp.
  101--112.

\bibitem{Lobel2014}
I.~Lobel, E.~D. Sadler, and L.~R. Varshney, ``Customer referral incentives and
  social media,'' \emph{Available at SSRN: http://ssrn.com/abstract=2520615},
  2014.

\bibitem{Karnik2012}
A.~Karnik and P.~Dayama, ``Optimal control of information epidemics,'' in
  \emph{Proc. 4th IEEE Int. Conf. Communication Systems and Networks
  (COMSNETS)}, 2012, pp. 1--7.

\bibitem{Dayama2012}
P.~Dayama, A.~Karnik, and Y.~Narahari, ``Optimal incentive timing strategies
  for product marketing on social networks,'' in \emph{Proc 11th Int. Conf.
  Autonomous Agents and Multiagent Systems-Volume 2}, 2012, pp. 703--710.

\bibitem{Kandhway2014}
K.~Kandhway and J.~Kuri, ``How to run a campaign: Optimal control of sis and
  sir information epidemics,'' \emph{Applied Mathematics and Computation}, vol.
  231, no.~0, pp. 79 -- 92, 2014.

\bibitem{Kandhway2014a}
------, ``Optimal control of information epidemics modeled as maki thompson
  rumors,'' \emph{Communications in Nonlinear Science and Numerical
  Simulation}, vol.~19, no.~12, pp. 4135--4147, 2014.

\bibitem{Kandhway2014b}
------, ``Campaigning in heterogeneous social networks: Optimal control of si
  information epidemics,'' \emph{IEEE/ACM Trans. Netw.}, vol.~PP, no.~99, pp.
  1--1, 2014.

\bibitem{Barrat2008}
A.~Barrat, M.~Barthelemy, and A.~Vespignani, \emph{Dynamical processes on
  complex networks}.\hskip 1em plus 0.5em minus 0.4em\relax Cambridge
  University Press, 2008.

\bibitem{Baxter2010}
G.~J. Baxter, S.~N. Dorogovtsev, A.~V. Goltsev, and J.~F. Mendes, ``Bootstrap
  percolation on complex networks,'' \emph{Phys. Rev. E}, vol.~82, no.~1, p.
  011103, 2010.

\bibitem{Dorogovtsev2010}
S.~N. Dorogovtsev, \emph{Lectures on complex networks}.\hskip 1em plus 0.5em
  minus 0.4em\relax Oxford University Press Oxford, 2010, vol.~24.

\bibitem{Granovetter1978}
M.~Granovetter, ``Threshold models of collective behavior,'' \emph{American
  journal of sociology}, pp. 1420--1443, 1978.

\bibitem{Macy1991}
M.~W. Macy, ``Chains of cooperation: Threshold effects in collective action,''
  \emph{American Sociological Review}, pp. 730--747, 1991.

\bibitem{Mahajan1991}
V.~Mahajan, E.~Muller, and F.~M. Bass, ``New product diffusion models in
  marketing: A review and directions for research,'' in \emph{Diffusion of
  technologies and social behavior}.\hskip 1em plus 0.5em minus 0.4em\relax
  Springer, 1991, pp. 125--177.

\bibitem{Berger2001}
E.~Berger, ``Dynamic monopolies of constant size,'' \emph{Journal of
  Combinatorial Theory, Series B}, vol.~83, no.~2, pp. 191--200, 2001.

\bibitem{Karmarkar1984}
N.~Karmarkar, ``{A New Polynomial-time Algorithm for Linear Programming},'' in
  \emph{Proc. 16th Annu. ACM Symp. Theory of Computing}, 1984, pp. 302--311.

\bibitem{snapnets}
J.~Leskovec and A.~Krevl, ``{SNAP Datasets}: {Stanford} large network dataset
  collection,'' \url{http://snap.stanford.edu/data}, Jun. 2014.

\bibitem{konect}
J.~Kunegis, ``{KONECT Datasets}: Koblenz network collection,''
  \url{http://konect.uni-koblenz.de/}, 2015.

\bibitem{moshe}
M.~Shaked and J.~G. Shanthikumar, \emph{Stochastic Orders}.\hskip 1em plus
  0.5em minus 0.4em\relax Springer, 2007.

\end{thebibliography}

\end{document}